\documentclass[a4paper,twocolumn]{revtex4}
\usepackage{amsmath}
\usepackage{amsfonts}
\usepackage{amssymb}
\usepackage{graphicx}
\usepackage{subscript}
\usepackage{siunitx}
\usepackage{xcolor}
\usepackage{gensymb}

\begin{document}

\title{Hierarchical self-assembly of anisotropic colloidal platelets}

\author{Carina Karner}
\email{carina.karner@univie.ac.at}
\affiliation{Faculty of Physics, University of Vienna, Boltzmanngasse 5, A-1090, Vienna, Austria}

\author{Christoph Dellago}
\affiliation{Faculty of Physics, University of Vienna, Boltzmanngasse 5, A-1090, Vienna, Austria}

\author{Emanuela Bianchi}
\email{emanuela.bianchi@tuwien.ac.at}
\affiliation{Institut f{\"u}r Theoretische Physik, TU Wien, Wiedner Hauptstra{\ss}e 8-10, A-1040 Wien, Austria }
\affiliation{CNR-ISC, Uos Sapienza, Piazzale A. Moro 2, 00185 Roma, Italy}

\date{\today}

\begin{abstract} 
Anisotropy at the level of the inter-particle interaction provides the particles with specific instructions for the self-assembly of target structures. The ability to synthesize non-spherical colloids, together with the possibility of controlling the particle bonding pattern $via$ suitably placed interaction sites, is nowadays enlarging the playfield for materials design. 
We consider a model of anisotropic colloidal platelets with regular rhombic shape and two attractive sites placed along adjacent edges and we run Monte Carlo simulations in two-dimensions to investigate the two-stage assembly of these units into clusters with well-defined symmetries and, subsequently, into extended lattices. Our focus is on how the site positioning and site-site attraction strength can be tuned to obtain micellar aggregates that are robust enough to successively undergo to a second-stage assembly from sparse clusters into a stable hexagonal lattice. 
\end{abstract}

\maketitle

\section{Introduction}
Nowadays, colloids of different shapes can be accurately synthesized within a vast range of symmetries, from complex convex units -- such as spheroidal colloidal molecules, rods or polyhedra -- to concave shapes -- such as multi-pods or bowl-shaped colloids~\cite{shapes_review,nanocrystals_review,colloidal_molecules_review}. 
On top of the shape-dependent anisotropic interaction, directionality in bonding is often sought after in order to fine tune the assembly of the particles into either finite clusters~\cite{Kraft_2012,Sacanna_2013,Smalyukh_2015,Azari_2017} or extended crystals~\cite{Granick_2012,Gang_2015,Lee_2016}. 
The combination of bonding sites or extended surface patches~\cite{patchyrevtheo,patchyrevexp,newreview} -- $i.e.$, regions of the particle surface where bonding occurs -- and non-spherical particle shapes imparts preferred bonding directions between the particles, thus creating a rich enthalpic $versus$ entropic competition to be taken advantage of for materials design~\cite{Murray_2013}. 

Finite clusters can serve, for instance, as prototypes of micro-robots to perform tasks at the micro-scale such as delivery and targeting. Recent and successful examples of anisotropic patchy particles assembling into finite clusters are colloidal asymmetric dumbbells that show a rotational propulsion under electric field~\cite{Wu_2015}, dielectric cubes with one metallic facet that reconfigure on demand~\cite{Velev_2017}, or Janus spheroids that act as encapsulating agents~\cite{Rickman_2015}. 

Self-assembled finite clusters can be also used as non-spherical building blocks for further -- hierarchical -- assembly into extended structures and thereby expand  the versatility of the original units \cite{Whitelam2015,Gruenwald2014,Chen2012,Morphew2018,Mehdi2016}. Colloidal molecules composed of different types of particles, for instance, can be designed to support the assembly of superstructures with target photonic or phononic properties~\cite{Zanjani_2019}.

In general, the combination of shape and bond anisotropy leads to an extraordinary control over the crystal structure: polyhedral nanoparticles covered with DNA surface ligands have been shown, for instance, to assemble into crystals fully determined by the size and the symmetry of the particles and by the length of the DNA-ligands~\cite{Mirkin_2015,Mirkin_2016}. DNA-based functionalization can also be achieved by creating suitably shaped frames for spherical nanoparticles: two-dimensional, square-like DNA frames with functionalized edges are able to tune the assembly of the resulting complex units from finite clusters (micelles as well as chain-like aggregates) to planar architectures~\cite{Gang_2016_2D}; while tetravalent DNA-cages are able to induce the assembly of isotropic nanoparticles into diamond crystals~\cite{Gang_2016_3D}.

Within this vast realm, colloidal platelets, $i.e.$, colloids with one dimension much smaller than the two others~\cite{shapes_review}, may show, $e. g.$,  interesting electronic properties as isolated fluorescent emitters~\cite{Efros_2011} or even form tilings with photonic properties~\cite{Watson_2017}. In general, colloidal platelets of different shapes can be realized both at the nanometer and the micron scale by, $e. g.$, folding long, single-stranded DNA molecules to create two-dimensional shapes~\cite{Rothemund_2006,Qian_2018}, by soft lithography~\cite{Lee_2016} or by making use of preferred growth mechanisms thus obtaining, $e.g.$, silica polygonal truncated pyramids, lanthanide fluoride nanocrystals or polymer-based platelets~\cite{Smalyukh_2015,Murray_2013,Manners_2017,Li_2019}. Directional bonding can then be added by, $e.g.$, covering the particle edges with ligands~\cite{Murray_2013} or immersing the platelets in a liquid crystal~\cite{Smalyukh_2015}. While the directional bonding constraints favor particle configurations that maximize the number of bonds, the particle shape favors edge-to-edge contacts; the possible competition between these two driving-agents can lead to self-assembled tilings with tunable properties~\cite{Karner_nanolett_2019}.

Here we consider hard colloidal platelets with a regular rhombic shape decorated with two mutually attractive patches arranged -- in different geometries -- along two adjacent particle edges. We define two classes of rhombi characterized by having the two patches enclosing either the big or the small angle and, for each of them, we move the patches either in a symmetric or asymmetric fashion along the particle edges. Thus, the resulting building blocks present a wide range of bonding patterns, the combination of which gives rise to different assembly products: linear or zigzag chains as well as micelles with three-, five- or six-fold symmetry. We first investigate the emerging assembly scenarios according to the patch positioning at different interaction energies. 
For most of the studied systems, we are able to trace a region in the parameter space where micelles are the prevalent assembly product.
Hence the second part of our paper investigates if and how systems of micelles with different geometries form a well-defined lattice via hierarchical assembly.

\section{Methods}

\subsection{Particle model}
We consider regular hard rhombi decorated with two attractive patches. The interaction potential between two of such hard particles, $i$ and $j$, is given by
\[ U(\vec{r}_{ij}, \Omega_{i}, \Omega_{j})  =
  \begin{cases}
    0     & \quad \text{if  $i$ and $j$ do not overlap}\\
    \infty  & \quad \text{if $i$ and $j$ do overlap}\\
  \end{cases}
\]
where $\vec{r}_{ij}$ is the center-to-center vector, while $\Omega_{i}$ and $\Omega_{j}$ are the particle orientations. Overlaps between rhombi are detected $via$ the separating axis theorem for convex polygons~\cite{Golshtein_1996}. 

The patch-patch interaction is a square-well attraction given by
\[ W(p_{ij})  =
  \begin{cases}
    - \epsilon     & \quad \text{if}\quad p_{ij}< 2r_{p}\\
    0 & \quad  \text{if}\quad p_{ij} \geq 2r_{p}, \\
  \end{cases}
\]
where $p_{ij}$ is the patch-patch distance, $2r_{p}$ is the patch diameter and $\epsilon$ denotes the patch interaction strength.

\begin{figure*}[ht]
\begin{center}
\includegraphics[width=0.8\textwidth]{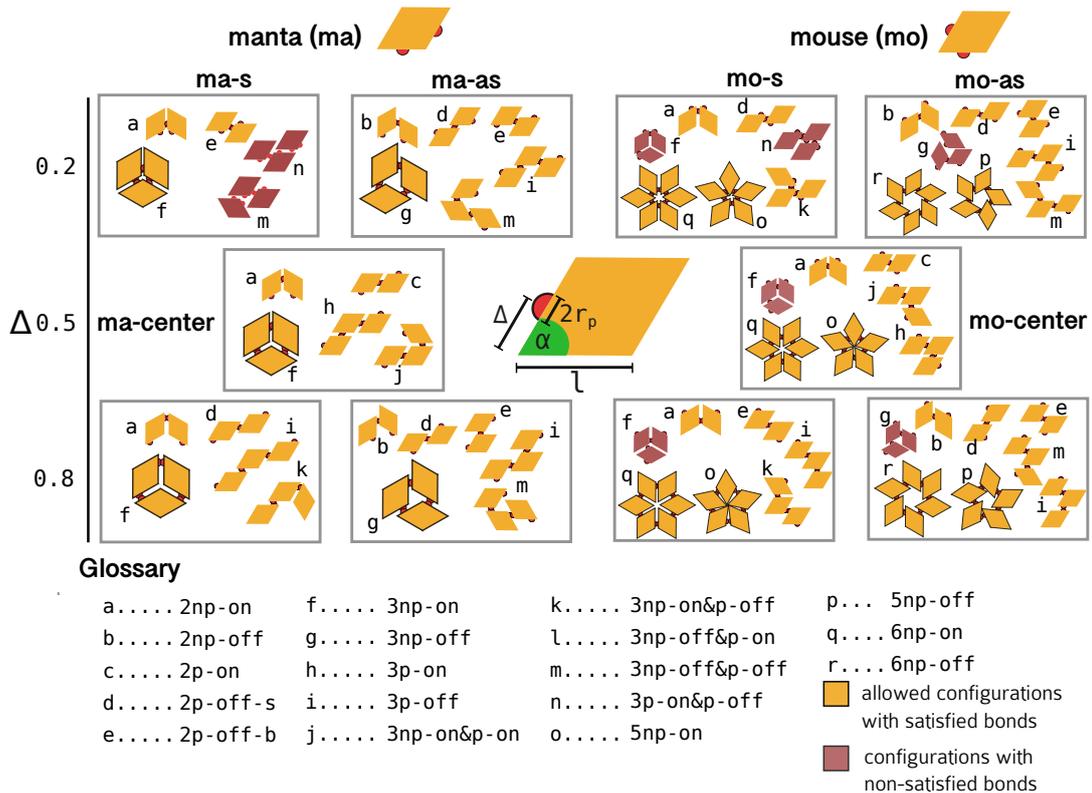}
\caption{Sketch of a patchy rhombus (in the center) and small cluster analysis for manta (ma) and mouse (mo) systems in the symmetric (s) and asymmetric (as) topologies. The columns correspond to different particle classes (ma, mo) and different topologies (s, as), as labeled; the rows correspond to different $\Delta$-values. At $\Delta=0.5$, the s- and as-topologies collapse to the center topologies, referred to as ma-center and mo-center. Clusters that satisfy bonding restrictions are colored in yellow and clusters that do not satisfy bonding restrictions and would yield non-satisfied bonds are colored in burgundy. Allowed micelles are enlarged and outlined in black. Cluster labels -- in the letters a-r -- are detailed in the glossary at the bottom. In the glossary, each configuration is characterized by the number of particles in the cluster (here from two up to six) and by the bond type: parallel (p) $vs$ non-parallel (np) as well as on-edge (on) $vs$ off-edge (off). Thus, dimers with on-edge non-parallel bonds (2np-on) are labeled with a, dimers with off-edge non-parallel bonds (2np-off) are labeled with b, dimers with on-edge parallel bonds (2p-on) are labeled with c, dimers with off-edge parallel bonds (2p-off) are labeled with either d (off-edge bonds closer to the small angle, 2p-off-s) or e (off-edge bonds closer to the big angle, 2p-off-b). The same logic applies to trimers and bigger clusters.
}
\label{fig:clusters}
\end{center} 
\end{figure*}

A patchy rhombi model of this kind was first detailed in~\cite{Whitelam2012}, with four attractive patches placed in the center of the edges. 
In this work, we focus on two-patch rhombi where the patches either enclose the big angle (manta rhombi, referred to as ``ma'')  or the small angle (mouse rhombi, referred to as ``mo"); see the particle sketches reported at the top of Fig.~\ref{fig:clusters}.

In either type of system, patches can be placed anywhere on the respective edges, resulting in an -- in principle -- infinite number of possible two-patch rhombi systems. To methodically characterize the patch positioning, we introduce two patch topologies. Patch topologies prescribe how to move the patches with respect to each other. In the symmetric/asymmetric (s/as) topology, patches are placed symmetrically/asymmetrically with respect to their enclosing vertex. Note that, within a specific topology, the relative distance $\Delta$ of one patch with respect to the enclosing vertex also determines  the position of the other patch. With these definitions a two-patch system is fully defined through its patch configuration (ma or mo), its topology (s or as) and its relative position on the edge ($\Delta$).  It is important to note that when patches are placed in the edge-center, $i.e.$, at $\Delta=0.5$, the s- and as-topology collapse into the same topology, referred to as center topology, so that the respective rhombi systems are referred to as ma- and mo-center systems. A summary of the used particle parameters can be found in Table~\ref{table:geom}.

\begin{table}[t]
\begin{center}
\begin{tabular}{ |l|l|l| } 
\hline
 \bf{parameter} & \bf{symbol} & \bf{value} \\
 \hline
 angle & $\alpha$ & $60\degree$ \\ 
 \hline
 side length & $l$ & 1.0 \\ 
 \hline
 patch radius & $r_{p}$  & 0.05 \\
 \hline
 interaction strength & $\epsilon$ & $[5.2 $, $10.2]$\\
 \hline
 patch position  & $\Delta$ & $[0.2,0.8]$ \\
 \hline
\end{tabular}
\caption{Particle parameters: the opening angle $\alpha$, the side length $l$ (which sets the unit of length), the patch size $2 r_p$, the patch-patch attraction strength $\epsilon$ (in unit of $k_BT$), the patch position $\Delta$, as labeled. Note that $\Delta$ refers to the relative placement of a patch on a rhombi edge. See Fig.~\ref{fig:clusters} for sketches.}
\label{table:geom}
\end{center}
\end{table}
\begin{table}[h]
    \begin{center}
    \begin{tabular}{|l|l|l|}
        \hline
        \bf{system parameter} &  \bf{symbol} & \bf{value} \\
        \hline
         area of simulation box & A & $1000 \cdot \sin{(60\degree)}$ \\
         \hline
         box width & $L_{x} $ & $\sqrt{1000}$ \\
         \hline
         box height & $L_{y}$ &  $\sqrt{1000}
         \cdot\sin{(60\degree)}$ \\
         \hline
         chemical potential eq. & $\mu_{eq}$ & 0.1 \\
         \hline
         chemical potential & $\mu^{*}$ & 0.3 \\
         \hline
         Boltzmann constant  & $k_{\rm B}$ & 1 \\
         \hline
         Temperature & $T$ & 0.1\\ 
         \hline
          Pressure & $P$ & $[5,100]$ \\ 
         \hline
    \end{tabular}
    \caption{System parameters (used in all simulations).}
    \label{table:system_param}
    \end{center}
\end{table}

\subsection{Simulation details}
For the first-stage assembly, we use grand canonical Monte Carlo (GCMC) simulations to model the adsorption of the self-assembling platelets on a surface~\cite{Whitelam2012}.  Together with single particle rotation/translation moves and particle insertions/deletions, we implement also cluster moves \cite{Whitelam2007, Whitelam2010} to avoid kinetic traps. We equilibrate the systems for $3\times10^5$ MC-sweeps at low a packing fraction ($i.e.$, at $\phi\approx 0.05$) with $\mu_{eq}$ and then we increase the chemical potential to $\mu^{*}$ to observe the assembly. We run the simulations for about  $\approx 3\times 10^6 - 5.0\times 10^6$ MC-sweeps before collecting statistics. In order to further improve statistics, we perform eight simulations runs per system and interaction strength. For the second-stage assembly -- from the aggregated clusters into super-lattices -- we run isobaric-isothermal Monte Carlo simulations (NPT-MC) with isotropic volume moves at different pressure values, $P$. The system parameters for all simulations are are given in Table~\ref{table:system_param}.

\section{Results}
In general, the investigated systems yield three different classes of self-assembly products: chains, loops and micelles, emerging from the specific interplay between steric constraints and patchiness, and characterized by specific bonding patterns. By construction, our patchy rhombi can form a maximum of two bonds per particle, one per edge. Each of such bonds can occur either when the rhombi are oriented parallel (p) to each other or when the rhombi are arranged in an arrowhead -- $i.e.$, non-parallel (np) -- configuration. Chains must have p-bonded but can also present np-bonds, loops require both p- and np-bonds, while micelles are minimal loops and consist of np-bonds only.

We find that micelles are mostly in competition with chain assemblies, as loops rarely form. While our previous publication detailed the properties of chain assemblies~\cite{Karner_jpcm_2019}, this work focuses on micelles.
First, within the framework of a small cluster analysis, we establish for which topologies and patch positions micelles can form (see Fig.~\ref{fig:clusters}). Subsequently, we determine for which ($\Delta$, $\epsilon$)-values micelles are the dominant assembly product in simulations (see Fig.s~\ref{fig:mamo_phases}, \ref{fig:mamo_as_phases} and \ref{fig:yields}). Finally, we compress the micelle assemblies with the highest yield and analyze the quality of the second-stage assembly products (see Fig.s~\ref{fig:npt_box}, \ref{fig:npt_open-box} and \ref{fig:npt_stars}).

\subsection{Small cluster analysis}
The small cluster analysis reported in Fig.~\ref{fig:clusters} allows us to discern which clusters fulfill the given bonding constraints.  Note that the bonding constraints themselves follow from the patch configuration (ma or mo), the patch topology (s or as) and the patch position ($\Delta$). In addition to p- and np-bonds, we distinguish between on-edge bonds (on) where the edges align and off-edge bonds (off) where the edges are offset with respect to each other.  In the case of the p-off bonds, we further specify if the bond is closer to the small (p-off-s) or to the  big angle (p-off-b) (see the small cluster analysis and the glossary of Fig.~\ref{fig:clusters} for sketches of bonding scenarios).

In ma-s, micelles consist of three np-on-bonded particles (also referred to as boxes) and they can form at all $\Delta$-values (see clusters labeled as f in the corresponding panels of Fig.~\ref{fig:clusters}). For $\Delta<0.5$ micelles are the only possible self-assembly product as bonding incompatibilities prevent dimers (see cluster e in the corresponding panel of Fig.~\ref{fig:clusters}) growing into chains and loops (see clusters n and m, respectively). In contrast, for $\Delta \geq 0.5$, chains and loops can form since the patch positioning does not disfavor anymore the formation of p- and mixed-bonded clusters of sizes bigger than two (see clusters h, i, j and k in the corresponding panels of Fig.~\ref{fig:clusters}).

In mo-s, micelles consist either of five or six np-on-bonded particles (referred to as 5- and 6-stars, respectively) and they can form at all $\Delta$-values (see clusters o and q in the corresponding panels of Fig.~\ref{fig:clusters}).  We note that, micelles of three particles (boxes) would have non-satisfied bonds, thus they are energetically disfavored with respect to the stars. For $\Delta<0.5$, three-particle clusters with p- and np-bonds are still allowed (see cluster k in the corresponding panel of Fig.~\ref{fig:clusters}), thus rendering chains a possible self-assembly product, albeit with a restricted configuration space. For $\Delta \geq 0.5$, also both p- and mixed-bonded clusters can form (see clusters i and m in the corresponding panel of  Fig.~\ref{fig:clusters}).

In both ma-as and mo-as,  there are no bonding restrictions for p-bonds, and hence chains, loops and micelles are allowed for all values of $\Delta$. Furthermore, micelles have pores in the center as particles can form np-bonds only off-edge. Therefore, the resulting clusters are referred to as open boxes in the case of ma-as (see clusters g  in the corresponding panels of Fig.~\ref{fig:clusters}) and open stars (see clusters p and r in the corresponding panels of Fig.~\ref{fig:clusters}). 
 
It is important to note that while in s-topologies the geometrical form of the assembled micelles is the same at all $\Delta$-values, in as-topologies the cluster geometry does change because of the $\Delta$-dependent opening and closing of the pores. While at $\Delta=0.5$ (corresponding to ma- and mo-center), the pores are completely closed yielding closed boxes and stars, the pores continuously open as $\Delta$ is moved off-center and -- as a result -- the rhombi corners protrude more and more. As in Ref.~\cite{Karner_nanolett_2019}, in the sticky limit, the side length of the pores is given by
\begin{equation}\label{eq:pores}
l_{\text{pore}}= \lvert l - 2\Delta \rvert.
\end{equation}
Once $l_{\text{pore}}$ is known, the area of the pores can be calculated for both triangular (ma-as systems) and hexagonal pores (mo-as systems). 

\subsection{First stage self assembly} 
Beyond the described bonding restrictions, we need to determine under which conditions the different aggregates emerge in many-body systems. For that, we run GCMC simulations and, at the end of the self-assembly process, we calculate yields of chains/loops and micelles.  

The yield of a cluster type is defined as the percentage of particles in clusters belonging to the selected cluster type. We define cluster types of interest according to the specific systems. In ma-systems  -- for both the s- and the as-topology -- we distinguish clusters of size $N<3$ (classified as liquid), three np-bonded particle loops (classified as boxes/open boxes) and non-box clusters with size $N\geq 3$ (corresponding to chains/loops). In mo-systems  -- both s and as --  we distinguish clusters with size $N<5$ (liquid),  five np-bonded particle loops (5-stars/open 5-stars), six np-bonded particle loops (6-stars/open 6-stars) and non-star clusters $N\geq 5$ (chains/loops). 

\begin{figure*}
\begin{center}
\includegraphics[width=\textwidth]{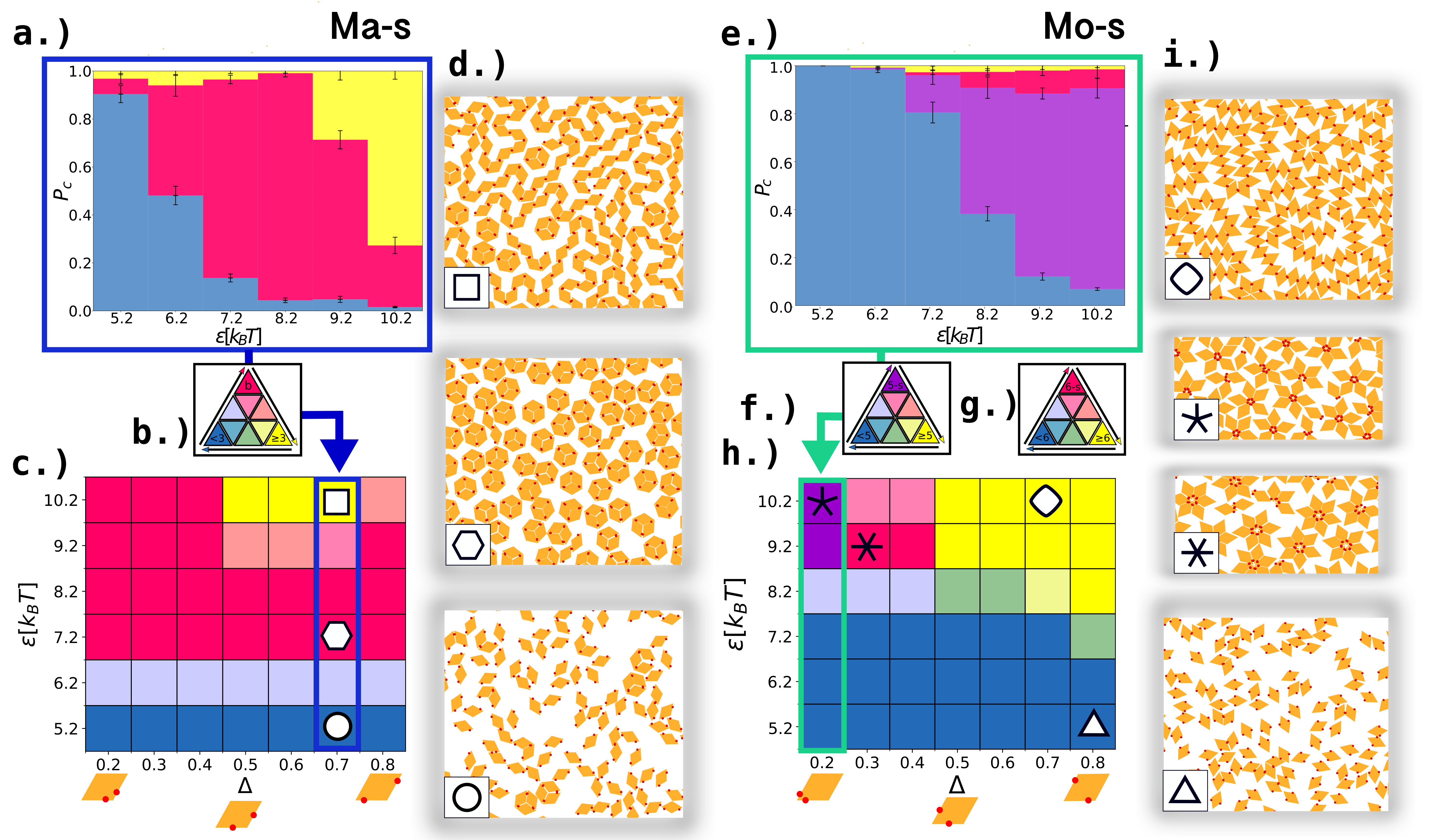}
\caption{Self-assembly products of symmetric manta (ma-s) and mouse (mo-s) systems. \textbf{a.)} Histogram of yields of different cluster types in the ma-s system at $\Delta=0.7$ and $\epsilon=  [5.2,10.2] k_{B}T$. Histograms for the other $\Delta$-values can be found in Fig.~\ref{fig:ma_s_histo} in Appendix \ref{section:all_yields}. Cluster types of interest are: clusters with size $N<3$ (liquid, blue), three np-bonded particle loops (boxes, pink) and clusters with $N\geq 3$ (chains/loops, yellow). The barycentric color triangle in \textbf{b.)} maps the distribution of cluster types for all ($\Delta, \epsilon$)-values to the heatmap in \textbf{c.)} (see Appendix \ref{section:barycentric} for the detailed description of the barycentric coordinates). \textbf{d.)} Snapshots of ma-s. From top to bottom: chains, loops and boxes at $\epsilon = 10.2 k_{B}T$, $\Delta = 0.7$, boxes at $\epsilon = 7.2 k_{B}T$, $\Delta = 0.7$, rhombi liquid at $\epsilon = 5.2 k_{B}T$, $\Delta = 0.7$.  \textbf{e.)} Histogram of yields of different cluster types in the mo-s system at $\Delta=0.2$ and $\epsilon = [5.2,10.2] k_{B}T$. Histograms for the other $\Delta$-values can be found in Fig.~\ref{fig:mo_s_histo} in Appendix \ref{section:all_yields}. Cluster types of interest are: clusters with $N<5$ (liquid, blue), five np-bonded particle loops (5-stars, purple), 6 np-bonded particle loops (6-stars, pink) and clusters with $N\geq 5$ (chains/loops,  yellow). The barycentric color triangles in \textbf{f.)}/\textbf{g.)} map the distribution of cluster types to the heatmap column/s for $\Delta = 0.2$/$[0.3,0.8]$ in \textbf{h.)} \textbf{i.)} Snapshots of mo-s. From top to bottom: chains, loops and stars at $\epsilon = 10.2 k_{B}T$, $\Delta = 0.7$, 5-stars at $\epsilon = 10.2 k_{B}T$, $\Delta = 0.2$, 6-stars at $\epsilon = 9.2 k_{B}T$, $\Delta = 0.3$, rhombi liquid at $\epsilon = 5.2 k_{B}T$, $\Delta = 0.8$.
}
\label{fig:mamo_phases}
\end{center} 
\end{figure*}

\begin{figure*}
\begin{center}
\includegraphics[width=\textwidth]{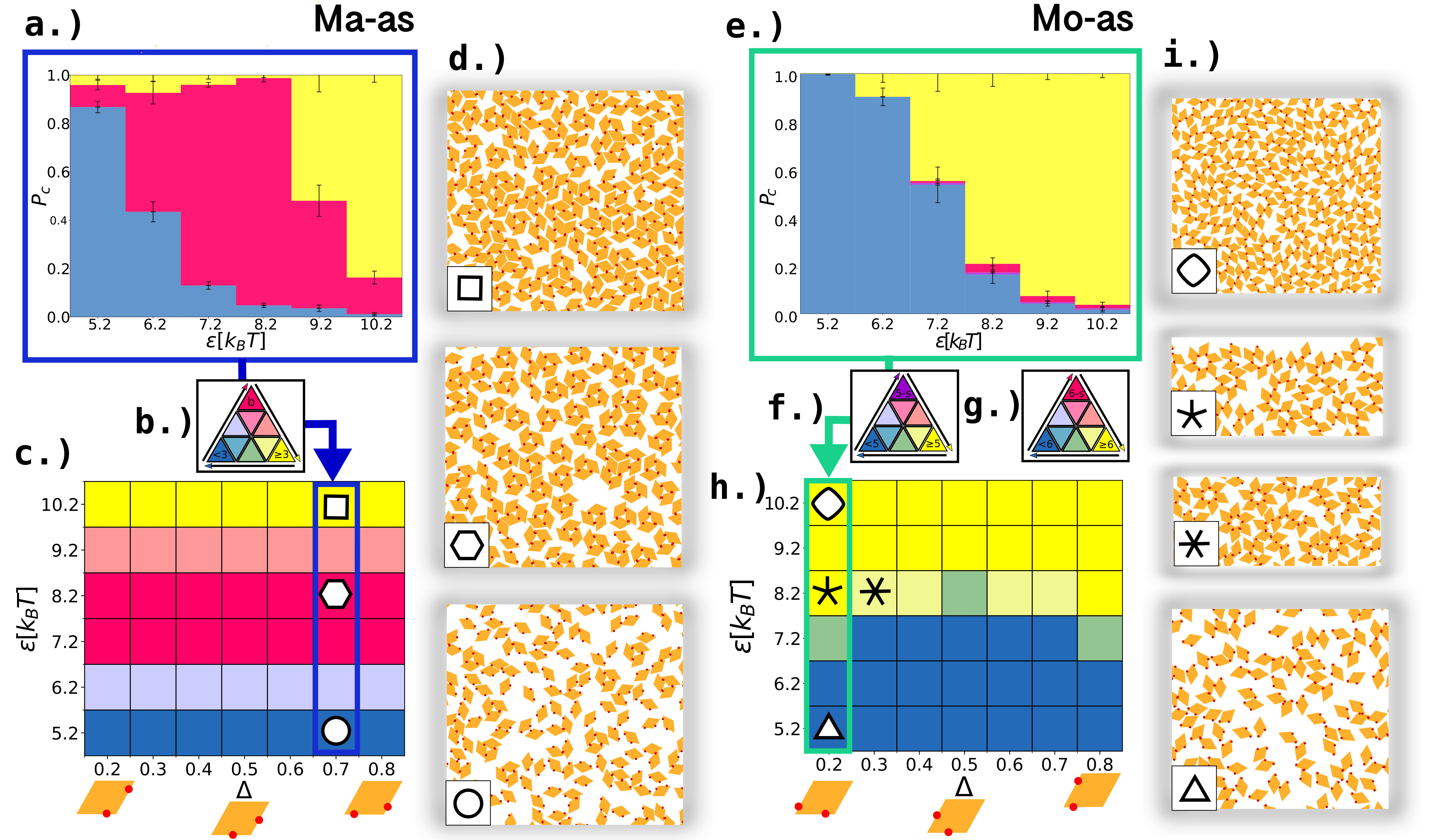}
\caption{Self-assembly products of asymmetric manta (ma-as) and mouse (mo-as) systems. \textbf{a.)} Histogram of yields of different cluster types in the ma-as system at $\Delta=0.7$ and $\epsilon = [5.2,10.2] k_{B}T$. Histograms for remaining $\Delta$ values can be found in Fig.~\ref{fig:ma_as_histo} in Appendix \ref{section:all_yields}. Cluster types of interest are: clusters with size $N<3$ (liquid, blue), three np-bonded particle loops (open boxes, pink) and clusters with $N\geq 3$ (chains/loops, yellow). The barycentric color triangle in \textbf{b.)} maps the distribution of cluster types for all ($\Delta, \epsilon$)-values to the heatmap in \textbf{c.)} (see Appendix \ref{section:barycentric} for the detailed description of the barycentric coordinates). \textbf{d.)} Snapshots of ma-as. From top to bottom: chains, loops and open boxes at $\epsilon = 10.2 k_{B}T$, $\Delta = 0.7$, open boxes at $\epsilon = 8.2 k_{B}T$, $\Delta = 0.7$, rhombi liquid at $\epsilon = 5.2 k_{B}T$, $\Delta = 0.7$. \textbf{e.)} Histogram of yields of different cluster types in the mo-as system at $\Delta=0.2$ and $\epsilon = [5.2,10.2] k_{B}T$. Histograms for the other $\Delta$-values can be found in Fig.~\ref{fig:mo_as_histo} in Appendix \ref{section:all_yields}. Cluster types of interest for $\Delta = 0.2 $ are: clusters with size $N<5$ (liquid, blue), five np-bonded particle loops (open 5-stars, purple), six np-bonded particle loops (open 6-stars, pink), and clusters with $N\geq 5$ (chains/loops,yellow). The barycentric color triangle in \textbf{f.)}/\textbf{g.)} maps the distribution of cluster types to the heatmap column/s for $\Delta = 0.2$/[0.3,0.8] in \textbf{h.)} \textbf{i.)} Snapshots of mo-as. From top to bottom: chains, loops and open stars at $\epsilon = 10.2 k_{B}T$, $\Delta = 0.2$, chains, loops and open stars at $\epsilon = 8.2 k_{B}T$, $\Delta = 0.2$, chains, loops and open stars at $\epsilon = 8.2 k_{B}T$, $\Delta = 0.3$, rhombi liquid at $\epsilon = 5.2 k_{B}T$, $\Delta = 0.2$.
}
\label{fig:mamo_as_phases}
\end{center} 
\end{figure*}

\begin{figure}
\begin{center}
\includegraphics[width=\columnwidth]{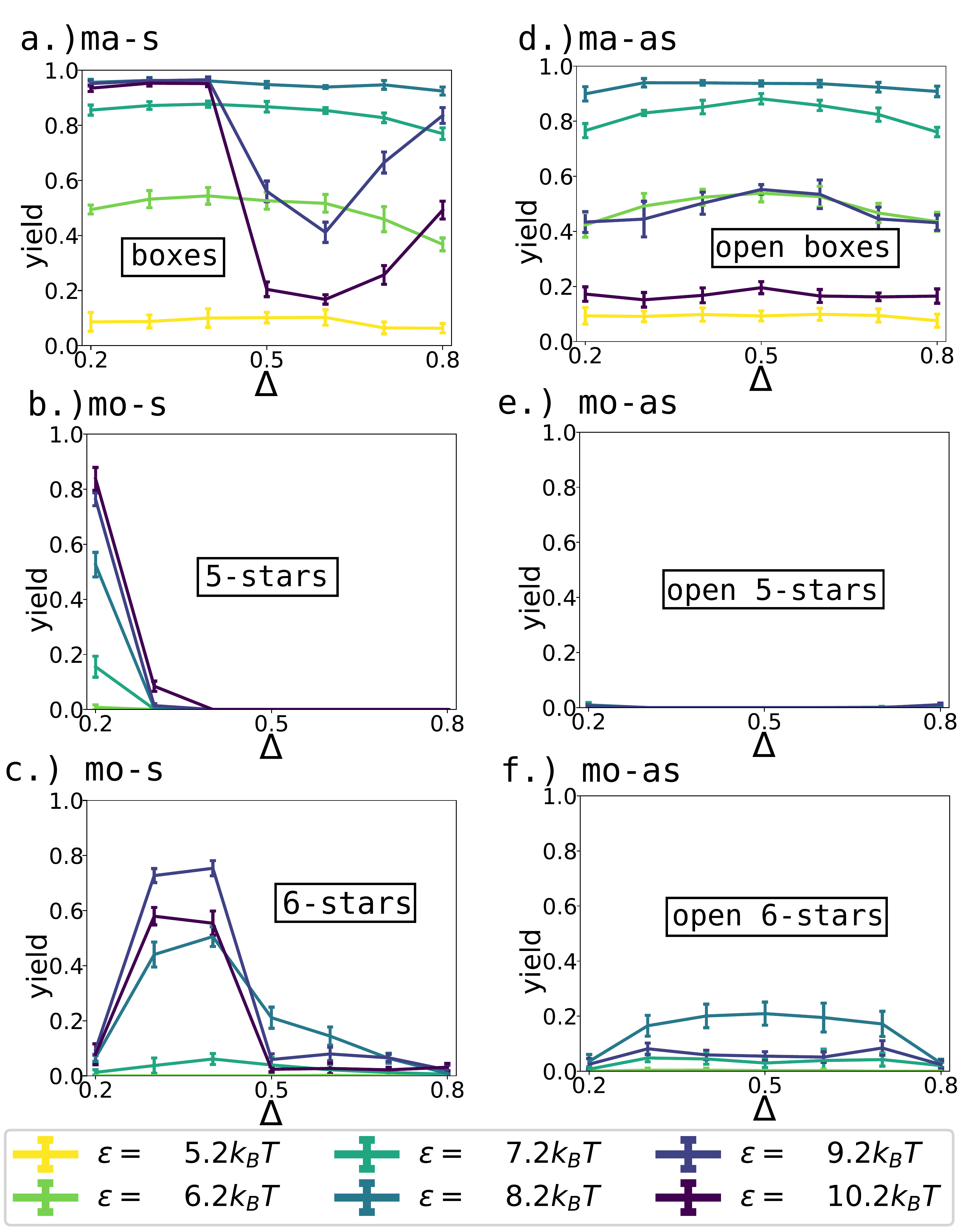}
\caption{Yields of micelles as a function of $\Delta$: \textbf{a.)}  boxes in ma-s systems. \textbf{b.)} 5-stars in mo-s systems.  \textbf{c.)} 6-stars in mo-s systems. \textbf{d.)} open boxes in ma-as systems. \textbf{e.)} open 5-stars in mo-as systems.  \textbf{f.)} open 6-stars in mo-as systems. Different curves within each panel denote yields at different $\epsilon$-values and are colored according to the legend at the bottom. Yields are averaged over eight simulation runs per system.}
\label{fig:yields}
\end{center} 
\end{figure}

The obtained yields are reported as a function of $\epsilon$ for each $\Delta$-value (see panels a. and e. of Fig.~\ref{fig:mamo_phases} and Fig.~\ref{fig:mamo_as_phases} for an example $\Delta$-value, for the full overview of $\Delta$-values see Fig.~\ref{fig:ma_s_histo}, \ref{fig:mo_s_histo}, \ref{fig:ma_as_histo} and \ref{fig:mo_as_histo} in Appendix \ref{section:all_yields}). Through a mapping to a barycentric coordinate system (see Appendix \ref{section:barycentric} for the mapping and calculation details), we obtain heatmaps for the four investigated systems (ma-s, mo-s, ma-as and mo-as) where at each ($\Delta$, $\epsilon$)-value we identify the predominant self-assembly scenario. The heatmaps reflect the relative dominance of the different cluster types: whenever one cluster type has a yield higher than $2/3$, the heatmap obtains the color of the respective triangle edge (blue for liquid, yellow for chains/loops, pink and purple for micelles) and we call this cluster type $2/3$-dominant; if no cluster type dominates by more than $2/3$, one of the mixed colors in the center of the triangle is adopted (see panels c. and h. of Fig.~\ref{fig:mamo_phases} and Fig.~\ref{fig:mamo_as_phases} for the heatmaps, see panels b., f. and g. in Fig.~\ref{fig:mamo_phases} and~\ref{fig:mamo_as_phases} for the respective barycentric triangles). Simulation snapshots are displayed in Fig.~\ref{fig:mamo_phases}d/i for ma-s/mo-s systems and in Fig.~\ref{fig:mamo_as_phases}d/i for  ma-as/mo-as systems. Micelle yields for all systems are given in Fig.~\ref{fig:yields}. 

We start our discussion with ma-s systems (Fig.~\ref{fig:mamo_phases}, panels from a. to d.). When $\epsilon \leq 6.2 k_{B}T$ such systems remain liquid over the whole $\Delta$-range. On increasing $\Delta$, the self-assembly process takes place: at $\epsilon=7.2 k_{B}T$ and $\epsilon=8.2 k_{B}T$, boxes are the $2/3$-dominant self-assembly product at all $\Delta$-values; while for $\epsilon \geq 9.2 k_{B}T$ there exists a range of $\Delta$-values where chains/loops prevail over the micelles (see Fig.~\ref{fig:mamo_phases}c). The yield of boxes increases monotonically with $\epsilon$ for all $\Delta$-values until $\epsilon = 9.2 k_{B}T$ (see Fig.~\ref{fig:yields}a). For $\epsilon \geq 9.2 k_{B}T$ and $\Delta < 0.5$, the yield of the boxes reaches values above $0.95$, while at $\Delta = 0.5$, we observe a sudden drop: for $\Delta \geq 0.5$ bonding restrictions allow chains/loops to form and compete with boxes and as soon as the interaction strength is $\epsilon\geq 9.2 k_{B}T$, chains/loops prevail over the boxes. The dip in the box-yield at ma-center becomes deeper on increasing $\epsilon$ and while for $\epsilon = 9.2 k_{B}T$ the yield recovers to $0.8$ at $\Delta=0.8$, for $\epsilon = 10.2k_{B}T$ the yield remains below $0.45$ at $\Delta=0.8$. Summarizing, in ma-s, boxes prevail over chains/loops over a wide range of ($\Delta$, $\epsilon$)-values.

In mo-s systems (Fig.~\ref{fig:mamo_phases}, panels from e. to i.), we observe characteristic self-assembly products only for $\epsilon \geq 8.2 k_{B}T$. At such high interaction strengths, micelles emerge only at $\Delta=0.2$ in the case of 5-stars and at $\Delta=0.3$ and $0.4$ in the case of 6-stars. For these $\Delta$-values, the star-yields rise monotonically with $\epsilon$ (see Fig.~\ref{fig:yields}b for the 5-star-yield and Fig.~\ref{fig:yields}c for the 6-star-yield), but only reach values above $2/3$ for $\epsilon \geq 9.2 k_{B}T$. The highest 5-star-yield is reached at $\epsilon = 10.2 k_{B}T$ and $\Delta=0.2$ with $0.838 \pm 0.04$. On increasing $\Delta$, the 5-star-yield falls to $0.084 \pm 0.018$ at $\Delta=0.3$ and becomes smaller than $0.025$ at $\Delta=0.5$. 
In contrast, 6-stars reach their highest yields at $\epsilon = 9.2 k_{B}T$, with the highest value $0.754 \pm 0.028$ at $\Delta=0.4$. On increasing $\Delta$,  the 6-star-yield drops to $\approx 0.2$ at mo-center and proceeds to drop gradually below $0.05$ at $\Delta=0.8$. With star-yields below $0.2$ for $\Delta \geq 0.5$, chains/loops prevail as dominant clusters at high $\epsilon$-values (see Fig.~\ref{fig:mamo_phases}h for phase boundaries).

Summarizing, in mo-s, stars have a significantly narrower $(\Delta, \epsilon)$-range of prevalence compared to boxes in ma-s (see Fig.~\ref{fig:mamo_phases}c for ma-s and Fig.~\ref{fig:mamo_phases}h for mo-s).
While boxes are $2/3$-dominant for all $\Delta$ at intermediate $\epsilon$-values, stars reach yields above $2/3$ only at high $\epsilon$-values and only for $\Delta<0.5$. 
At high $\epsilon$ and for $\Delta\geq 0.5$ extended p-bonded chains become available to mo-s and are formed at a higher rate. 

In ma-as systems (Fig.~\ref{fig:mamo_as_phases}, panels from a. to d.), the yield of micelles (open boxes, in this case) rises monotonously with $\epsilon$: open boxes become $2/3$-prevalent for intermediate interaction strengths ($i.e.$, at $\epsilon = 7.2 k_{B}T$ and  $8.2 k_{B}T $) over the whole $\Delta$-range, with the highest yields -- above $0.9$ -- achieved at $\epsilon = 8.2 k_{B}T$ (see Fig.~\ref{fig:mamo_as_phases}c for the heatmap and Fig.~\ref{fig:yields}d for numerical values of yields). At high interaction strengths, $i.e.$, for $\epsilon \geq 9.2 k_{B}T$, the open box yield drops for all $\Delta$,  first to $0.4$ at $\epsilon = 9.2 k_{B}T$ and then to $0.2$ at $\epsilon = 10.2k_{B}T$. The drop in the micelle-yield is due to the emergence of chains, that can form at all $\Delta$-values and prevail over the open boxes when $\epsilon \geq 9.2 k_{B}T$. 

In mo-as systems (Fig.~\ref{fig:mamo_as_phases}, panels from e. to h.), micelles (open stars, in this case) do not become dominant at any $(\Delta, \epsilon)$-value. In particular, the yield of open 5-stars stays below $0.03$ for all $\Delta$-values and does not increase with $\epsilon$, meaning that these micelles are always negligible compared to chains/loops (see Fig.~\ref{fig:yields}e). In contrast, the yield of open 6-stars does increase with $\epsilon$ for all but the most extreme $\Delta$-values ($i.e.$, $\Delta=0.2$ and $0.8$); nonetheless, as the maximum yield stays below $0.25$ (see Fig.~\ref{fig:yields}f), also these micelles are negligible over the whole $\Delta$-range. While for $\epsilon \leq 7.2 k_{B}T$, the systems are mostly liquid, for $\epsilon \geq 7.2 k_{B}T$, chains become the most dominant cluster type at all $\Delta$-values. 

Concluding, micelles in as-topologies have a smaller window of prevalence compared to micelles in s-topologies because in as-topologies extended p- and mixed bonded chains are allowed at all $\Delta$-values and they dominate over micelles for high interaction strengths. In the following, we describe the properties of the observed micelles, mostly focusing on their packing properties. For the detailed characterization of the chain/loops assemblies we refer the reader to Ref.~\cite{Karner_jpcm_2019}. 

\subsection{Second-stage assembly}

To further explore the versatility of patchy rhombi as building blocks, we compress configurations with high micelle yields and thereby induce a second-stage assembly process. We use the configurations as they formed within simulations, including all not fully formed and misshaped clusters and monomers. This enables us to study the efficiency of the hierarchical assembly process, from two-patch rhombi to micelle lattices and show viable routes for material synthesis. Within our computer simulations, the second-stage self-assembly is modeled with NPT-MC simulations and the pressure range is $P=5-100$. We selected systems with a high percentage of micelles, $i.e.$, with yields higher than $0.75$. While there are $(\Delta, \epsilon)$-values for which boxes, open boxes and stars reach yields above $0.75$, maximum yields of open stars stay below $0.25$. Therefore we exclude them from our second-stage assembly investigation and focus on boxes, open boxes and stars. Note that boxes, open boxes and stars are hard particles: boxes are convex as they are regular hexagons; open boxes are non-convex, with their protrusions growing more prominent as $\Delta$ moves towards off-center-values; 5-stars and 6-stars are non-convex.

\begin{figure*}[ht]
\begin{center}
\includegraphics[width=0.9\textwidth]{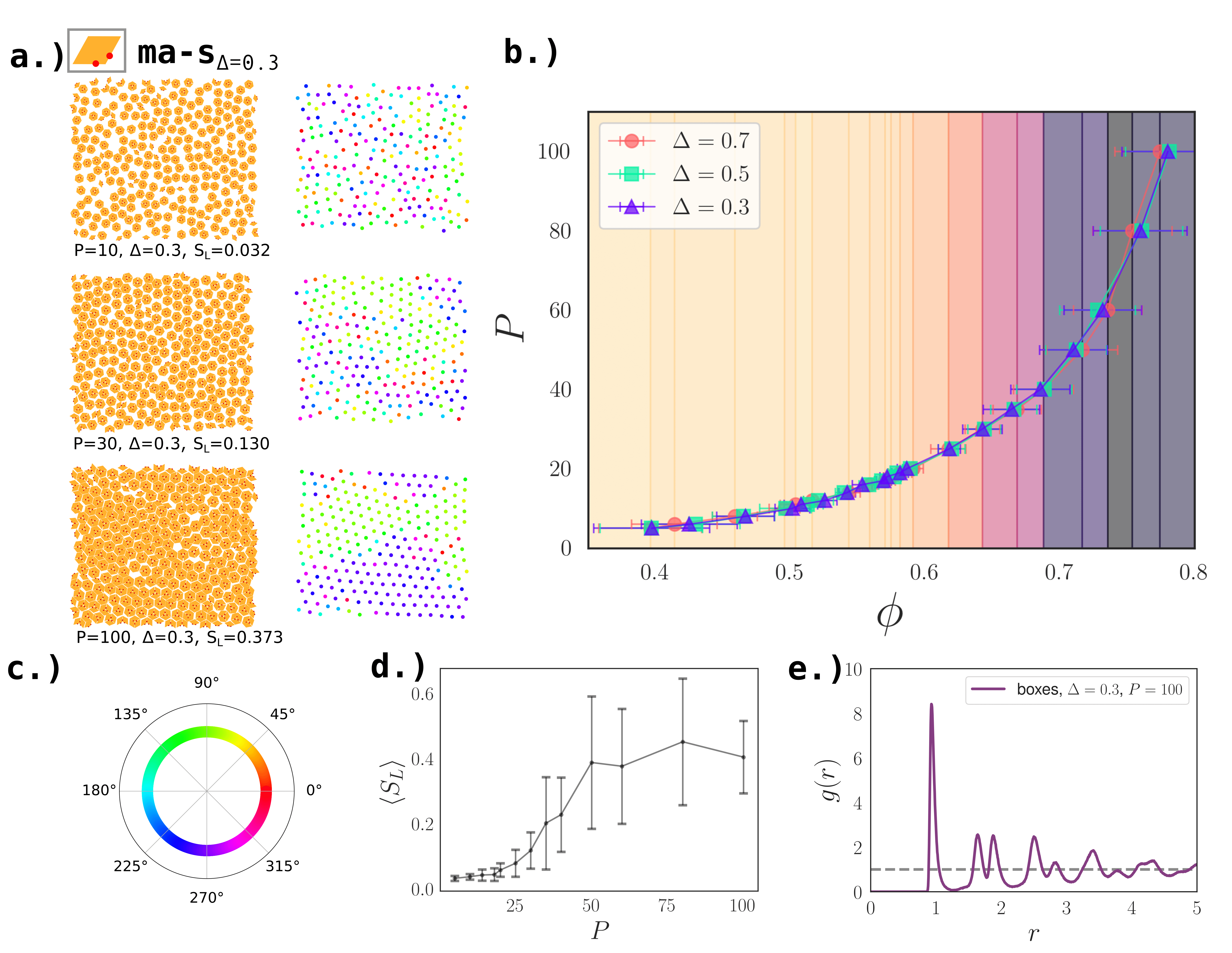}
\caption{Second-stage assembly of boxes at $\epsilon = 8.2 k_{B}T$ for symmetric manta systems (ma-s). \textbf{a.)} Left: snapshots of the system with $\Delta=0.3$ at $P=[10,30,100]$ (from top to bottom). Right: centers of mass of boxes of the respective snapshots colored according to the hexatic order parameter, $\Psi$, defined in Eq.~\ref{eq:psi} (see the color wheel in panel c). \textbf{b.)} Eos for box systems at $\Delta = [0.3,0.5,0.7]$, where $\phi$ denotes the packing fraction defined in Eq.~\ref{eq:phi}. \textbf{c.)} Color wheel denoting the values of  $\Psi$ for the center of mass snapshots in panel a. \textbf{d.)} Average fraction of largest hexagonal domain $\langle S_{L} \rangle$ -- defined in the text -- as function of $P$; note that the average is taken over all $\Delta$-values; note that the background of the eos-plot in panel a is colored according to $\langle S_{L} \rangle$. Error bars denote the standard deviation and their wide extent is due to relatively small system sizes. \textbf{e.)} The radial distribution function $g(r)$ at $\Delta = 0.3$ and $P=100$.
}
\label{fig:npt_box}
\end{center} 
\end{figure*}

\begin{figure*}
\begin{center}
\includegraphics[width=0.9\textwidth]{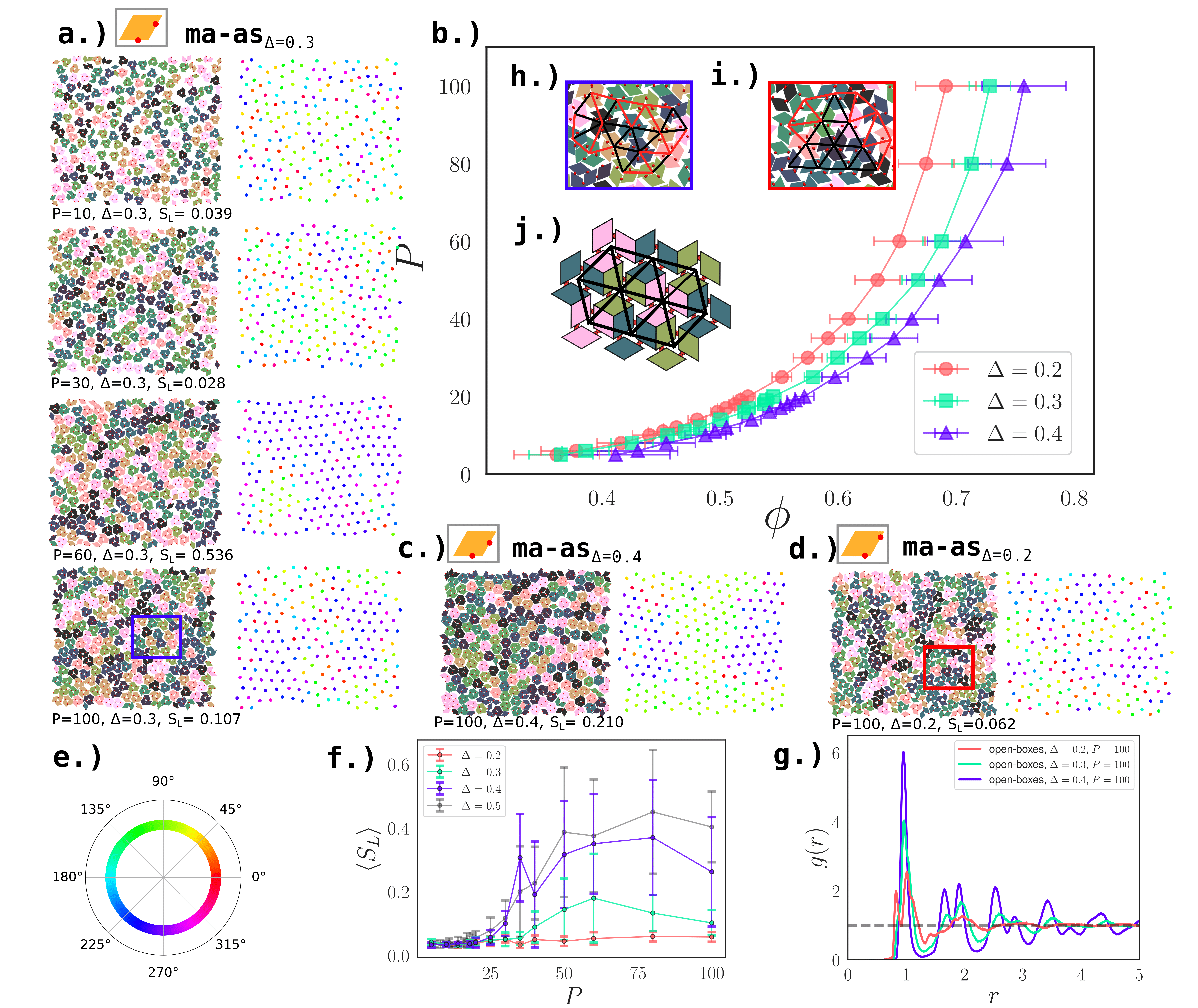}
\caption{Second-stage assembly of open boxes at $\epsilon = 8.2 k_{B}T$ for asymmetric manta systems (ma-as). \textbf{a.)} Left: snapshots of the system with $\Delta = 0.3$ at $P=[10,30,60,100]$ (from top to bottom); clusters are colored to facilitate visual distinction. Right: centers of mass of open boxes of the respective snapshots colored according to the hexatic order parameter, $\Psi$, defined in Eq.~\ref{eq:psi} (see the color wheel in panel e). \textbf{b.)} Eos for open box systems at $\Delta = [0.2,0.3,0.4]$, where $\phi$ is the packing fraction defined in Eq.~\ref{eq:phi}; \textbf{c.)} Left: snapshot of the system at $\Delta=0.4$ and $P=100$. Right: centers of mass of open boxes of the respective snapshot colored according to $\Psi$. \textbf{d.)} Left: snapshot of the system at $\Delta=0.2$ and P=$100$. Right: centers of mass of open boxes of the respective snapshot colored according to $\Psi$. \textbf{e.)} Color wheel for the values of $\Psi$ for the center of mass snapshots in panels a, c and d. \textbf{f.)} Average fraction of largest hexagonal domain $\langle S_{L} \rangle$  -- defined in the text --  as function of $P$ for $\Delta=[0.2,0.3,0.4,0.5]$, as labeled. The error bars denote the standard deviation and are wide due to relatively small system sizes. \textbf{g.)} The radial distribution function $g(r)$ at $P=60$ and $\Delta = [0.2, 0.3, 0.4]$, as labeled. \textbf{g.)} The radial distribution function $g(r)$ at $\Delta = [0.2, 0.3, 0.4]$ and $P=100$, as labeled. \textbf{h.)} Magnified view of blue rectangle of snapshot of ma-as$_\Delta=0.3$, P=100. 
Black lines connect boxes at hexagonal lattice positions; red lines connect boxes that are shifted due to defects. \textbf{i.)} Magnified view of red rectangle of ma-as$_\Delta=0.2$. Black lines connect boxes at hexagonal lattice positions; red lines connect boxes that are shifted due to defects. \textbf{j.)} Sketch of the perfect hexagonal open box tiling with triangular pores (ot-tiling) as observed in local regions in simulation snapshots.
}
\label{fig:npt_open-box}
\end{center} 
\end{figure*}

\begin{figure*}[ht]
\begin{center}
\includegraphics[width=0.9\textwidth]{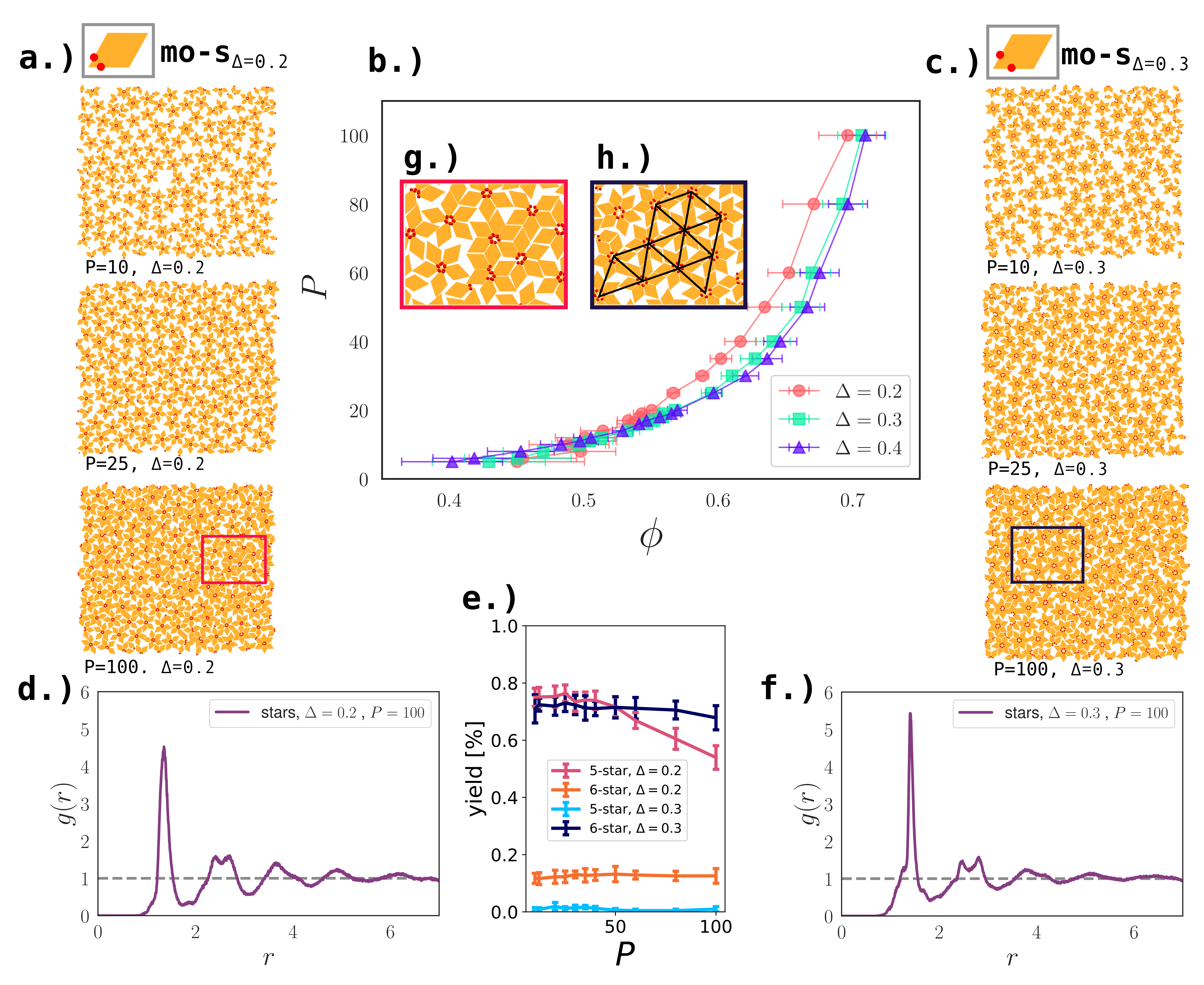}
\caption{Second-stage assembly of 5-stars and 6-stars at $\epsilon = 9.2 k_{B}T$ for symmetric mouse systems (mo-s).\textbf{a.)} Snapshots of the system at $\Delta = 0.2$ at $P=[10,25,100]$ (from top to bottom), where 5-stars are predominant.  \textbf{b.)} Eos for star systems at $\Delta = [0.2,0.3,0.4]$, where $\phi$ is the packing fraction defined in Eq.~\ref{eq:phi}.\textbf{c.)} Snapshots of the system at $\Delta = 0.3$ at $P=[10,25,100]$ (from top to bottom), where 6-stars are predominant. \textbf{d.)} The radial distribution function $g(r)$ for the 5-star system at $\Delta = 0.2$ and $P=100$.  \textbf{e.)} The yield of 5-stars and 6-stars at $\Delta = 0.2$ and $\Delta = 0.3$ as function of $P$, as labeled.  \textbf{f.)} The radial distribution function $g(r)$ for the 6-star system at $\Delta = 0.3$ and  $P=100$.
\textbf{g.)}. The red inset is a magnified view of 5-star neighbourhoods at $\Delta=0.2$ and $P=100$ \textbf{h.)}. The dark blue inset is magnified view of a 6-star neighborhoods at $\Delta=0.3$ and $P=100$. The black lines connect the star centers
and highlight the hexagonal order of this neighbourhood. 
}
\label{fig:npt_stars}
\end{center} 
\end{figure*}

Upon increasing pressure, micelles pack and tend to form crystalline domains with local hexagonal order. We define the packing fraction as
\begin{equation}\label{eq:phi}
\phi = N_p \frac{A_p}{A},    
\end{equation}
where  $N_{p}$ is the total number of particles, $A_p$ is the area of a rhombic platelet and $A$ is the total area of the simulation box. The equations of state (eos) for the different systems at selected $\Delta$-values are reported in panel b of Fig.~\ref{fig:npt_box} (ma-s), Fig.~\ref{fig:npt_open-box} (ma-as) and Fig.~\ref{fig:npt_stars} (mo-s).  
To characterize the emergence of long range order, we calculate the radial distribution function, $g(r)$, where $r$ is the center-to-center distance between two micelles. At this point we note that the side length of our rhombi is $l=1$. 
Additionally, we measure the local order with the hexatic order parameter, that is given for every micelle $j$ as 
\begin{equation}\label{eq:psi}
\Psi_{j} = \frac{1}{6}\sum_{k=1}^{N_{n}} \exp(i6\Theta_{j,k}),
\end{equation}
where $N_{n}$ is the number of all next neighbors and $\Theta_{j,k}$ is the relative bond angle between $j$ and its neighbor micelle $k$, with respective to a chosen simulation box vector. 
To determine all next neighbours, we calculate a distance histogram including the first 12 neighbours for every micelle and we define next neighbours as all micelles within the first minimum of this histogram.
Finally, to quantify the transition between isotropic to hexatic, 
we calculate the average fraction of particles in the largest hexagonal domain $\langle S_{L} \rangle$. Note that we consider two micelles, $i$ and $j$ as part of the same domain if the difference of their hexatic order parameter $d\Psi_{ij}<15\degree$.

\textbf{Box assembly.}
To study the second-stage assembly of boxes, we select ma-s systems with $\Delta = [0.3,0.5,0.7]$ and $\epsilon = 8.2 k_{B}T$.  The box-yields for these systems are $0.963 \pm 0.010$ at $\Delta =0.3$, $0.948 \pm 0.011$ at $\Delta=0.5$ and $0.947\pm0.016$ at $\Delta=0.7$. 

Upon increasing pressure, boxes assemble into a lattice with hexatic/hexagonal order at all selected $\Delta$-values. On the left hand side of Fig.~\ref{fig:npt_box}a, we display simulation snapshots for boxes with $\Delta = 0.3$ at different pressures, while on the right hand side we show the same snapshots with the box centers colored according to their respective $\Psi$ (see Fig.~\ref{fig:npt_box}c for $\Psi$-color wheel). We note that, while at $P=10$, the system of boxes is liquid, at $P\approx 20$ the system becomes hexatic, $i.e.$, with local hexagonal order but no long range order. As $P$ increases further, long range order becomes stronger. At $P=100$, remaining single rhombi induce grain boundaries into the otherwise perfectly hexagonally ordered domains. We confirm the hexatic/hexagonal order at $P=100$ with the radial distribution function (see Fig.~\ref{fig:npt_box}e). Further, we calculate  $\langle S_{L} \rangle$ as function of the pressure (see Fig.~\ref{fig:npt_box}d) to visualize the growth of ordered domains we colored the background of the eos-curves (see Fig.~\ref{fig:npt_box}b). The same scenario is observed at all selected $\Delta$-values: from $\Delta = 0.3$ to $\Delta = 0.7$ the equations of state collapse onto one curve (see Fig.~\ref{fig:npt_box}b). The maximum packing at the highest pressure $P=100$ is $\phi= 0.793 \pm 0.006$ for all $\Delta$-values. Therefore we conclude that for these boxes, the specific $\Delta$-value does not influence the second-stage assembly.

It is important to note that although significant parts of the systems are ordered hexagonally at high pressures, the small system sizes render it impossible to determine whether the system is hexagonal on a long range scale. See \cite{Krauth2011} for detailed studies on the liquid-hexatic-hexagonal phase transitions.

\textbf{Open box assembly.}
For the assembly of open boxes into super structures, we choose ma-as systems at $\epsilon=8.2k_{B}T$ and $\Delta=[0.2,0.3,0.4]$. Note that at $\Delta=0.5$ ma-as collapses to ma-center, thus forming boxes (already discussed in Fig.~\ref{fig:npt_box}).  As $\Delta$ moves towards off-center values, pores in the boxes open up and become bigger and bigger, while at the same time the corners protrude more and more. As a consequence, the eos-curves gradually shift to lower packing fractions at intermediate and high $P$-values (see Fig.~\ref{fig:npt_open-box}b). While boxes at $\Delta = 0.5$ and $P=100$ yield a packing of $0.793 \pm 0.006$, for open boxes at $\Delta=0.3$ the packing is $0.728 \pm 0.017$, and at $\Delta=0.2$ it is $0.691 \pm 0.025$. 

As with the boxes, we also measure the positional order with $g(r)$, $\langle S_{L} \rangle$ and $\Psi$ for the open boxes. The peak pattern of the $g(r)$ shows that hexagonal order is pronounced at $\Delta=0.4$, significantly deteriorates already at $\Delta=0.3$ and ceases completely at $\Delta=0.2$ (see Fig.~\ref{fig:npt_open-box}g). Similarly, $\langle S_{L}\rangle$ indicates smaller ordered domains for $\Delta=0.3$ and $\Delta = 0.2$ (note that due to large error-bars -- that are a result of the relatively small system sizes -- we can only report trends for $\langle S_{L} \rangle$). The progressive reduction of the hexatic/hexagonal order with increasing $\Delta$ can also be directly observed in the simulation snapshots reported in Fig.~\ref{fig:npt_open-box}a, c and d. In conclusion, in ma-as the hexatic/hexagonal order gets destroyed as $\Delta$ shifts towards off-center values.

\textbf{Star assembly.}
For the second-stage assembly of stars, we select systems at $\epsilon = 9.2 k_{B}T$ with $\Delta=0.2$ (5-stars) and $\Delta = [0.3,0.4]$ (6-stars). The star yields of these initial configurations are $0.838 \pm 0.041$ for 5-stars at $\Delta=0.2$, $0.727 \pm 0.025$ for 6-stars at $\Delta=0.3$ and $0.754 \pm 0.028$ for 6-stars $\Delta=0.4$.
It is important to note that these yields from CGMC simulations decay for all NPT-MC runs, even at low pressures $P$ (see Fig.~\ref{fig:npt_stars}e.).

The eos-curve for 5-stars (red dots in Fig.~\ref{fig:npt_stars}b) lies slightly above the 6-star curves for all intermediate and high $P$-values, indicating that 5-stars pack less close at the same $P$ with respect to 6-stars (green squares and blue triangles in Fig.~\ref{fig:npt_stars}b). The packing fraction for 5-stars at $P=100$ is $0.696 \pm 0.0214$, while for 6-stars at $\Delta=0.3$ it is $0.706\pm 0.018$. 

The analysis of the $g(r)$  at $P=100$ (panels d and f of Fig.~\ref{fig:npt_stars}) suggests that in both star systems there exists a residual hexagonal order. However, the resulting structures can not be described as hexagonal lattices due to the many non-star clusters present already in the initial configuration -- as the star-yields are low compared to those observed for the boxes.  Simulation snapshots in Fig.~\ref{fig:npt_stars}a and Fig.~\ref{fig:npt_stars}c show, that these clusters often have the structure of non-completed or misshaped stars. These ``broken stars" induce defects to an extent that destroys local and, subsequently, long range order.

Moreover, the 5-star-yield reduces on increasing pressure (see Fig.~\ref{fig:npt_stars}e): while the yield of 6-stars at $\Delta=0.3$ only slightly decreases with $P$, the 5-star yield at $\Delta=0.2$ decreases from $0.75 \pm 0.03$ at $P=10$ to $0.54\pm0.04$ at $P=100$.  This is due to the fact that, as the pressure is increased, the void where a sixth star would fit, starts to fill up.
The result are conglomerates of destroyed 5-stars (see Fig.~\ref{fig:npt_stars}b, red inset). Together with already present grain boundaries, these conglomerates completely break the hexatic/hexagonal lattice order.

In contrast, for 6-stars, areas with high local ordering suggest an open hexagonal lattice with triangular pores independent of $\Delta$ (see Fig.~\ref{fig:npt_stars}b, dark blue inset).

\section*{Conclusion}
In this paper, we investigate the assembly of patchy colloidal platelets with a regular rhombic shape. We consider a simple, purely two-dimensional model consisting of rhombi decorated with two bonding patches, arranged in different ways along adjacent particle edges.
Depending on patch arrangement and interaction strength we find micelles (minimal loops) or chains to be the prevalent self-assembly product. 

\begin{figure}
\begin{center}
\includegraphics[width=0.5\textwidth]{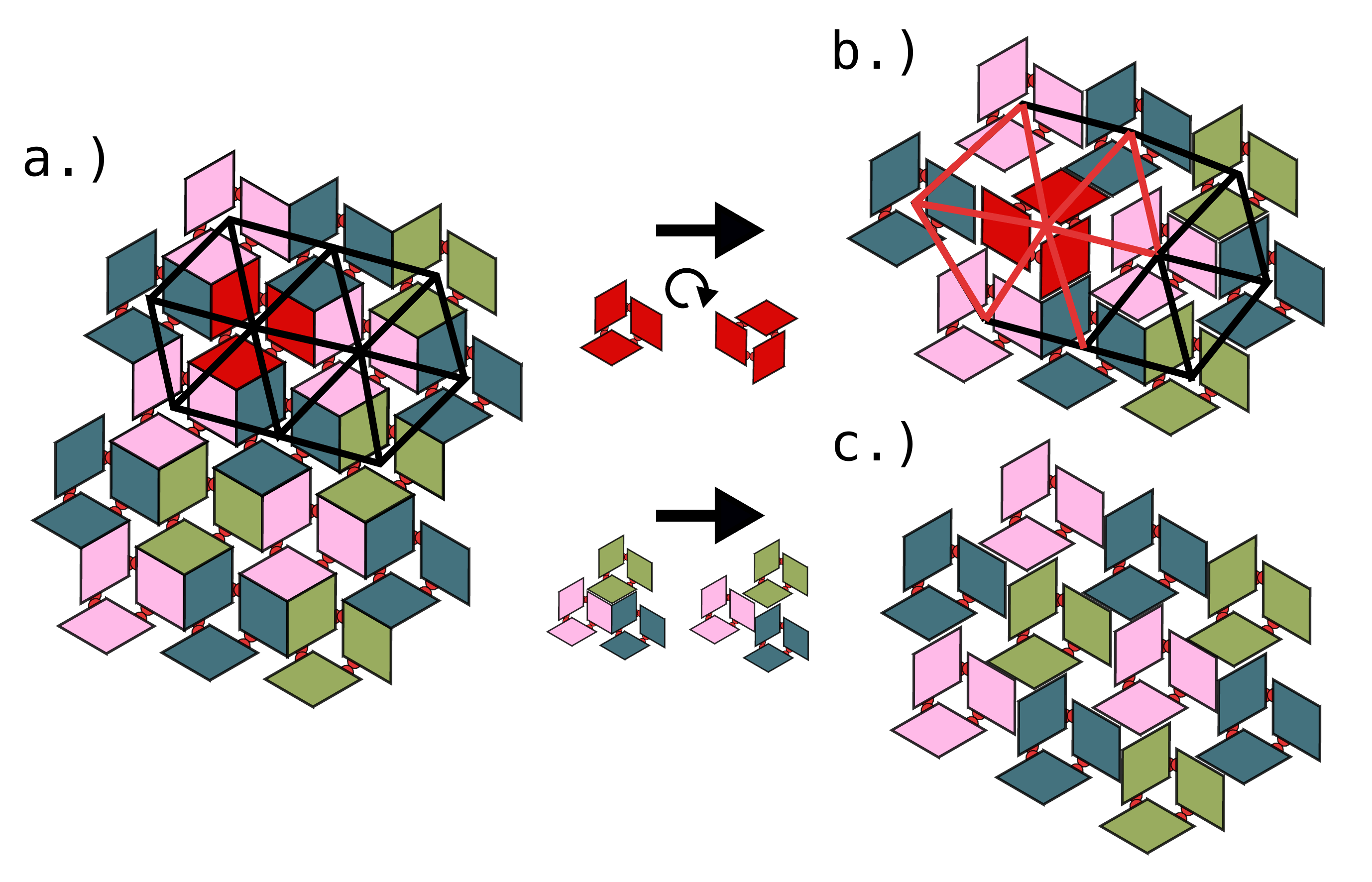}
\caption{Tilings of open boxes where the clusters are depicted in different colors to facilitate visual distinction. \textbf{a.)} Perfect hexagonal tiling with triangular pores, referred to as open triangular tiling (ot-tiling). The black lines connect the centers of neighboring boxes and highlight the hexagonal order of the tiling. \textbf{b.)} Representation of a lattice defect: the red open box is rotated by $60\degree$ with respect to the perfect ot-tiling in panel a. Due to the rotation, the rhombi of the red open box are aligned parallel with their neighboring rhombi belonging to a different box. The hexagonal lattice is thus distorted: while neighboring boxes in correct lattice positions are still connected by black lines, boxes that had to shift due to the orientational defect are connected with red lines. \textbf{c.)} Perfect hexagonal tiling of open boxes with triangular and hexagonal pores, referred to as open triangular-hexagonal tiling (oth-tiling). This tiling -- also known as kagome lattice -- was found in Ref.~\cite{Karner_nanolett_2019} in systems of patchy rhombi with four patches.}
\label{fig:box_tilings}
\end{center} 
\end{figure}

While an earlier work studied the properties of chains \cite{Karner_jpcm_2019}, this work focuses on micelles. 
In general, we distinguish two types of micelles: boxes, that emerge for systems where the patches enclose the big angle (manta), and stars that are formed by systems where the patches enclose the small angle (mouse). 

Boxes and stars can be stabilized for systems where both patches are distributed symmetrically at a distance less than $0.5$ ($\Delta<0.5$) from their enclosing angle. 
For systems with $\Delta\geq 0.5$, micelles and chains compete at high interaction strengths.

In systems, where patches are distributed asymmetrically - $i.e$ the distance of one patch to the enclosing angle is $\Delta$, while the distance of the second patch is $1-\Delta$ -- stars and boxes have a pore in the center and we call them open boxes and open stars.
We note that while for large as well as small $\Delta$-values the pore size is maximal, the pore closes off completely for $\Delta=0.5$ and symmetric and asymmetric patch distributions are identical. 
We find that, while open boxes are prevalent at intermediate interaction strength, open stars never reach a yield over $0.25$.
At high interaction strength chains prevail for all $\Delta$-values in both, manta and mouse systems. 

In the second part of the paper we compress the systems of boxes, open boxes and stars and characterize the emerging lattices. Of course, the yield of the first structures plays an important role in the second-stage assembly process~\cite{Gruenwald2014}, so we compress only those systems where micelle yield is sufficiently high.

While boxes yield a hexagonal lattice independent of $\Delta$, open boxes self-assemble into a hexagonal lattice with a $\Delta$-dependent long range order. As $\Delta$ becomes more extreme, gaps and pores within the lattice grow larger, while the long range order decays. 
In contrast, the self-assembly product of stars displays no long range hexagonal order for any $\Delta$-value. 

We note that, in box systems defects in the hexagonal order are exclusively due to either vacancies or not fully formed boxes. In open box and star systems, on the other hand, defects inherent to the particle shape arise.

For what concerns open boxes, the best hexagonal order would be achieved by the lattice reported in Fig.~\ref{fig:box_tilings}a. We refer to this lattice as open triangular tiling (ot-tiling) as it has triangular pores, whose size depend on $\Delta$. In the ot-tiling, all rhombi are aligned in a non-parallel fashion. Although such an alignment is able to incorporate the protruding corners while maximizing edge-to-edge contact, other on-edge alignments are possible and can poison the local order. As a consequence, in simulations, we observe orientational point defects due to the fact that rhombi belonging to different boxes align in a parallel way, as shown in Fig~\ref{fig:box_tilings}b. While for closed boxes this kind of parallel alignment is commensurable with the closed-packed hexagonal lattice, in open box systems, lattice distorting defects are introduced and the more extreme $\Delta$ is, the more the lattice is distorted. While at $\Delta=0.4$ and $\Delta=0.3$ open boxes still achieve long range order and high packing, at $\Delta=0.2$ the orientation defects destroy the lattice completely: no long range order is observed and a lower packing is reached. 

We note that, the ot-tiling is reminiscent of the perfect kagome lattice observed in  Ref.~\cite{Karner_nanolett_2019}, that is a hexagonal lattice with two kinds of pores: triangular and hexagonal ones. Both pore sizes are dependent on $\Delta$, with the pores becoming larger at more extreme $\Delta$-values. We refer to such a tiling as open triangular-hexagonal tiling (oth-tiling) and we sketch it in Fig.~\ref{fig:box_tilings}c. To map the ot-tiling to the oth-tiling, the on-edge alignment of the unbonded edges -- $i. e.$, those edges that are not decorated with patches -- must shift to an off-edge alignment. This shift would open up the hexagonal pores of the oth-tiling. Clearly, the ot-tiling prevails over the oth-tiling because hard particles prefer arrangements that maximise edge-to-edge contacts. To stabilize a defect-free kagome lattice, neighboring boxes must be forced into an off-edge alignment. One strategy to reach this goal might involve the presence of patches in the outer perimeter of the open boxes in positions that stabilize the off-set edge contacts. The specific topology supporting the oth-tiling in Ref.~\cite{Karner_nanolett_2019} can in fact be interpreted as a four patch extension of the ma-as topology that forms the open boxes. 

Analogue to open boxes, we speculate that, if the star yield was high enough, an ot-tiling could be the assembly product of the 6-stars systems (see Fig.~\ref{fig:defects}c in Appendix \ref{section:design}). In this case, the distortion from such an hexagonal perfect lattice would be less extreme with respect to the open boxes as neighboring stars have no choice but to align parallel, thus leading to line defects (see Fig.~\ref{fig:defects}d in Appendix \ref{section:design}). Local hexagonal arrangements and line defects can be observed in simulations; however the hexagonal order does not prevail there as stars are not stable under compression. To enhance the stability of stars -- and thus the quality of the second-stage assembly products -- parallel bonds between the rhombi should be disfavored as it is this type of bonds that allow broken stars and favor chains (see Fig.~\ref{fig:broken_stars}a in Appendix \ref{section:design}). A strategy to disfavor parallel bonds might consist in having different kinds of patches, where different types of patches attract each other (see Fig.~\ref{fig:broken_stars}b in Appendix \ref{section:design}).

\section*{Acknowledgements}
CK and EB acknowledge support from the Austrian Science Fund (FWF) under Proj. No. Y-1163-N27. Computation time at the Vienna Scientific Cluster (VSC) is also gratefully acknowledged. VMD was used to generate the simulation snapshots~\cite{vmd}.

\appendix
\renewcommand\thefigure{\thesection.\arabic{figure}}    
\setcounter{figure}{0}   

\section{Barycentric coordinates}
\label{section:barycentric}
With the help of barycentric coordinates it is possible visualize mixtures of three components $p_{l}, p_{m}, p_{c}$ that fulfill $p_{l} + p_{m} + p_{c} = 1$. This requirement is given in our case where $p_{i}$, with $(i=l,m,c)$ are the yields 
of the different cluster types. 
For ma-systems, $p_{l}$ is the yield of clusters with $N<3$, $p_{m}$ is 
the micelle yield, and $p_{c}$ is the yield of non-micelle clusters with $N\geq 3$. 
To use barycentric coordinates in mo-systems, we need to map the yield of four cluster types to the three components $p_{i}$.
As 5-stars and 6-stars are dominant at different $\Delta$-values we decided on the following mapping: 
for $\Delta=0.2$, $p_l$ corresponds to the yield of clusters with $N<5$, $p_{m}$ is the yield of micelle clusters with $N=5$, while all non-5-star clusters with $N\geq 5$ map to $p_{c}$;
in contrast for all $\Delta>0.2$, $p_{l}$ represents all clusters with $N<6$, $p_{m}$ is the yield of micelle clusters with $N=6$, and $p_{c}$ encompasses all non-6-star clusters with $N\geq 6$.

With these definitions, $(p_{l},p_{m},p_{c})$ are mapped onto a equilateral triangle. 
Each triangle vertex corresponds to the case where one $p_{i} = 1$, and the others are 0, whereas more mixed yields are represented by points within the triangle. 
If we set the Cartesian triangle edge points to be 
$x_{l} = (0,0)$, $x_{c} = (1,0)$ and $x_{m} = (1/2, \sqrt(3)/2)$,
the barycentric vector $q_{t}$ is given by
\begin{equation}
q_{t} = (\frac{1}{2}\cdot (2 p_{c} + p_{m}), \frac{\sqrt(3)}{2} \cdot p_{m})
\end{equation}
The equilateral triangle subsequently split into 9 equilateral triangles and the colors of the edge triangles blue, pink and yellow denote the case where one one cluster type has a yield higher than $2/3$.
In the case of mixed systems where no cluster type is $2/3$ dominant, the set $(p_{l}$,$p_{m}$,$p_{c})$ maps to points closer to the triangle center and the adopted color corresponds to the respective inside triangles.

\section{Histograms of yields}
\label{section:all_yields}
\setcounter{figure}{0}   

\begin{figure*}
\begin{center}
\includegraphics[width=0.8\textwidth]{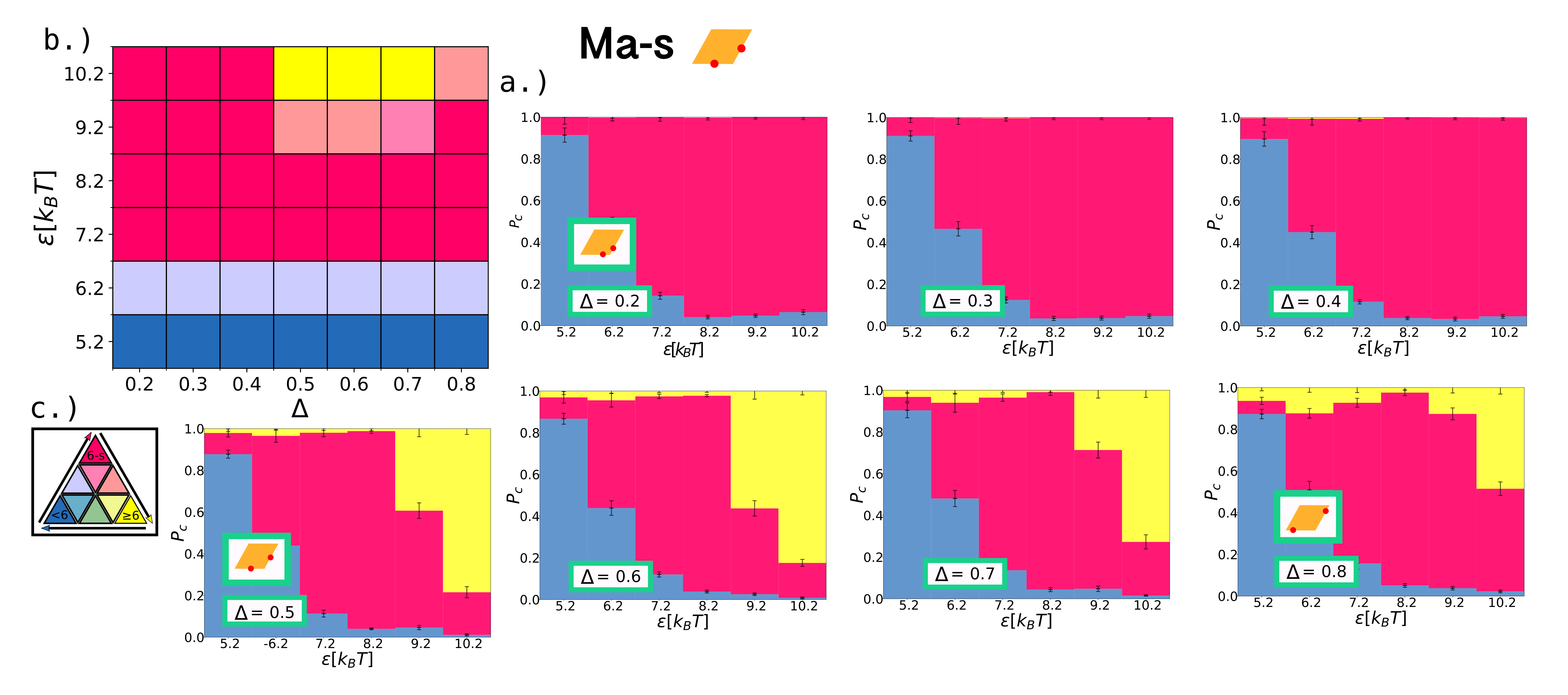}
\caption{Yield histograms and heatmaps for symmetric manta systems (ma-s). \textbf{a.)} histograms of yields for the cluster types liquid (blue), boxes (pink) and chain/loops (yellow). See the main paper for the definitions of the cluster types. There is one histogram plot for each patch position $\Delta$, the x-axis of each of these histograms denotes the interaction strength $\epsilon$. \textbf{b.)} The heatmap summarizes all yield histograms through mapping the yield distributions for each ($\Delta$, $\epsilon$) to a barycentric color triangle. \textbf{c.)} The barycentric color triangle for ma-s.}
\label{fig:ma_s_histo}
\end{center} 
\end{figure*}

\begin{figure*}
\begin{center}
\includegraphics[width=0.8\textwidth]{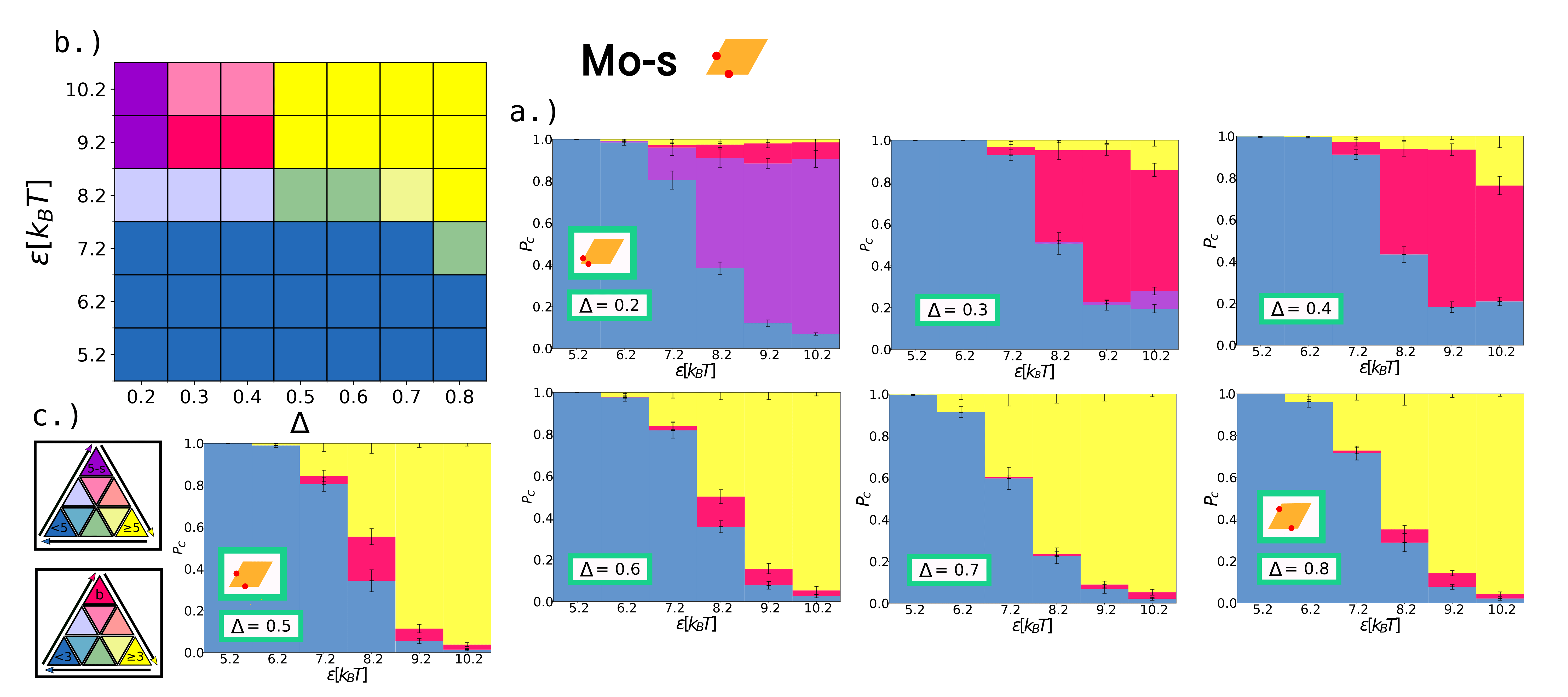}
\caption{Yield histograms and heatmaps for symmetric mouse systems (mo-s). \textbf{a.)} histograms of yields for the cluster types liquid (blue), 6-stars (pink), 5-stars(purple) and chain/loops (yellow). See the main paper for the definitions of the cluster types. There is one histogram plot for each patch position $\Delta$, the x-axis of each of these histograms denotes the interaction strength $\epsilon$. \textbf{b.)} The heatmap summarizes all yield histograms through mapping the yield distributions for each ($\Delta$, $\epsilon$) to a barycentric color triangle. \textbf{c.)} The barycentric color triangles for mo-s, where the upper triangle is used only to map the histogram of $\Delta = 0.2$, and the lower triangle triangle is used for all other $\Delta$-values.}
\label{fig:mo_s_histo}
\end{center} 
\end{figure*}

\begin{figure*}
\begin{center}
\includegraphics[width=0.8\textwidth]{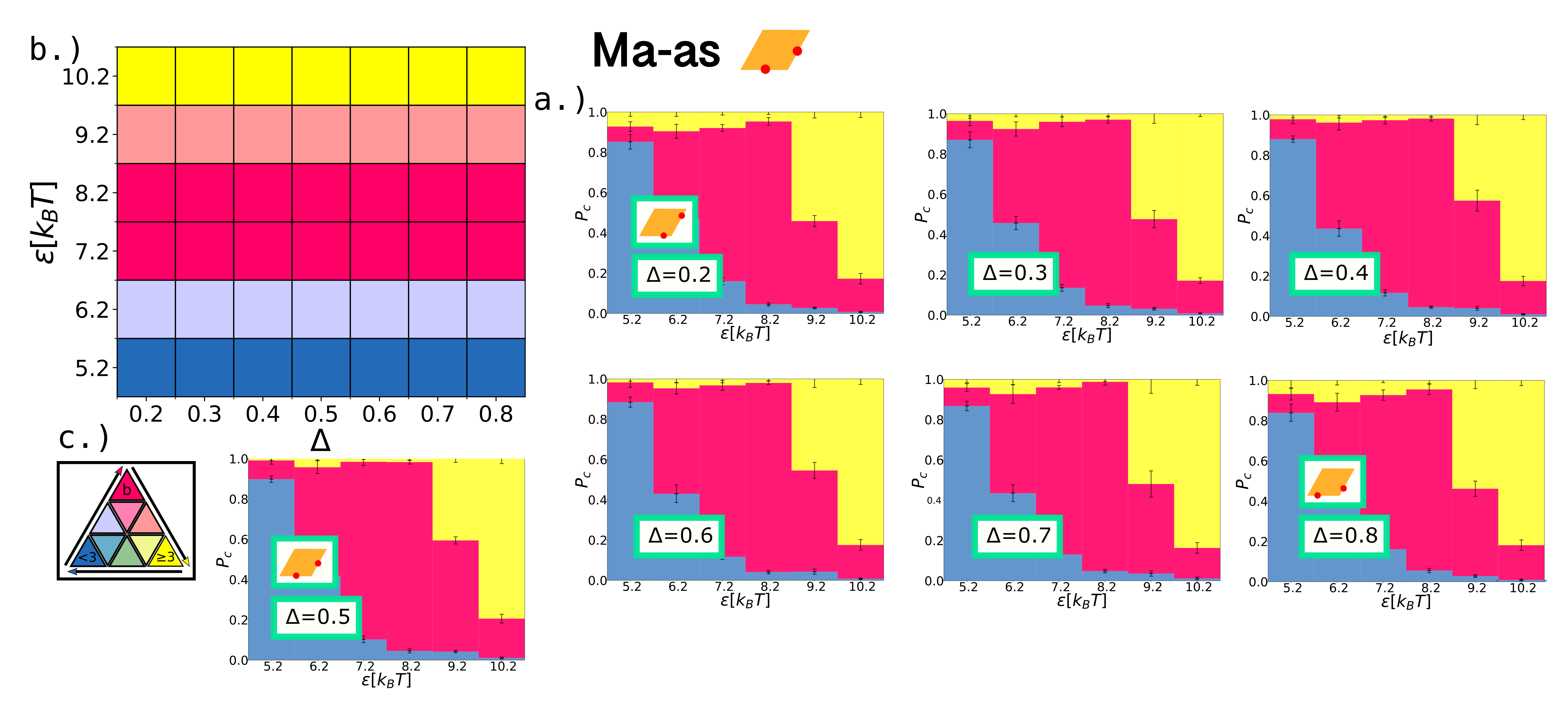}
\caption{Yield histograms and heatmaps for asymmetric manta systems (ma-as). \textbf{a.)} histograms of yields for the cluster types liquid (blue), open boxes (pink) and chains/loops (yellow). See the main paper for the definitions of the cluster types. There is one histogram plot for each patch position $\Delta$, the x-axis of each of these histograms denotes the interaction strength $\epsilon$. \textbf{b.)} The heatmap summarizes all yield histograms through mapping the yield distributions for each ($\Delta$, $\epsilon$) to a barycentric color triangle. \textbf{c.)} The barycentric color triangle for ma-as.}
\label{fig:ma_as_histo}
\end{center} 
\end{figure*}

\begin{figure*}
\begin{center}
\includegraphics[width=0.8\textwidth]{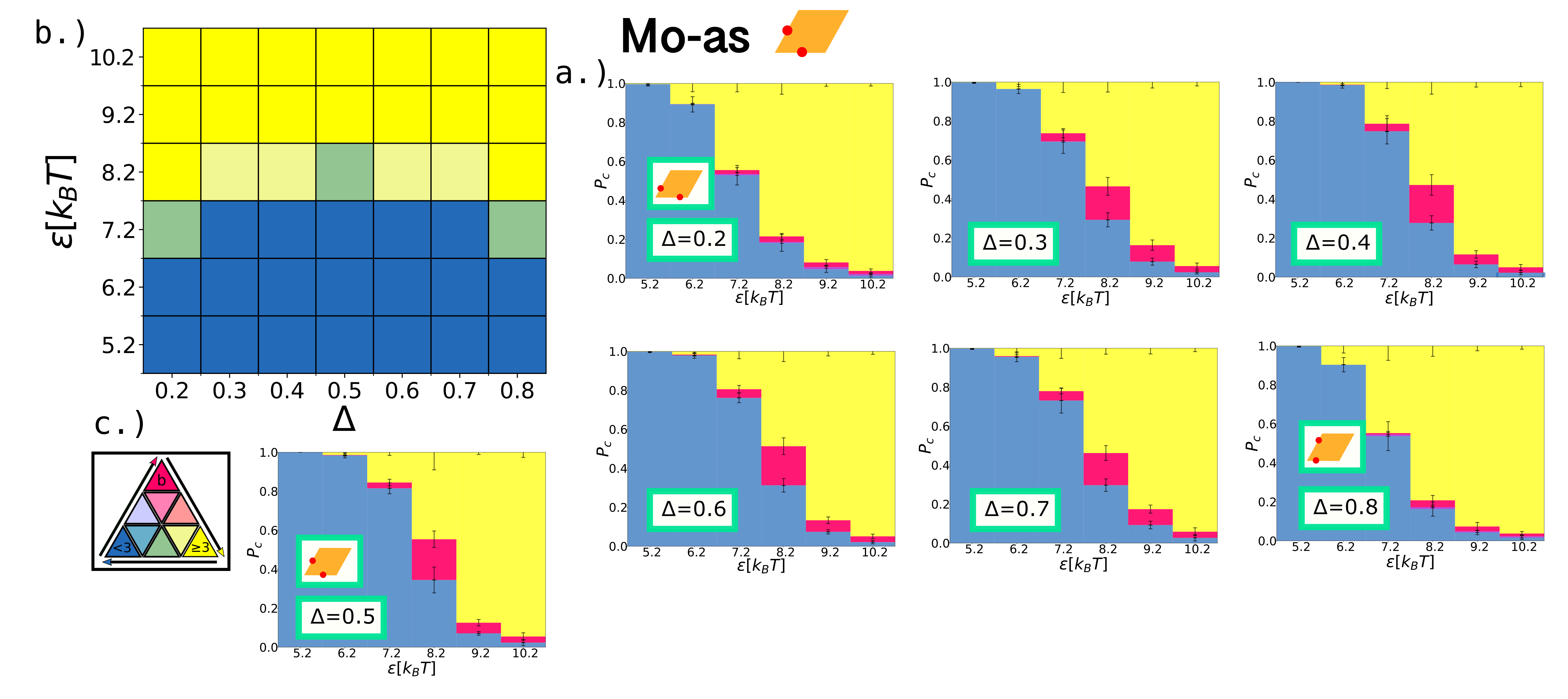}
\caption{Yield histograms and heatmaps for asymmetric mouse systems (mo-s). \textbf{a.)} histograms of yields for the cluster types liquid (blue), open 5-stars (purple), open 6-stars (pink) and chain/loops (yellow). See the main paper for the definitions of the cluster types. There is one histogram plot for each patch position $\Delta$, x-axis of each of these histograms denotes the interaction strength $\epsilon$. \textbf{b.)} The heatmap summarizes all yield histograms through mapping the yield distributions for each ($\Delta$, $\epsilon$) to a barycentric color triangle. \textbf{c.)} The barycentric color triangle for mo-as.}
\label{fig:mo_as_histo}
\end{center} 
\end{figure*}

We calculate the yields of the liquid state, the micelles and the chains/loops for all particle classes (manta (ma) and mouse (mo)) and all topologies (symmetric (s) and asymmetric (as)). See the main paper for definitions of the liquid state, the micelles, and the chain/loop clusters. The yields for all pairs ($\Delta$, $\epsilon$) are summarized in the histograms in Fig.~\ref{fig:ma_s_histo}a, \ref{fig:mo_s_histo}a, \ref{fig:mo_as_histo}a and \ref{fig:mo_as_histo}a.
With barycentric coordinates (see Sec.\ref{section:barycentric}) we map the yields to heat maps displayed in Fig.~\ref{fig:ma_s_histo}b, \ref{fig:mo_s_histo}b, \ref{fig:mo_as_histo}b and \ref{fig:mo_as_histo}b as well as in Fig.~2c/h and 3c/h in the main paper. The respective barycentric color triangles are displayed in Fig.~\ref{fig:ma_s_histo}c, \ref{fig:mo_s_histo}c, \ref{fig:ma_as_histo}c and \ref{fig:mo_as_histo}c and Fig.~2b/f/g and 3b/f/g in the main paper.

\section{Design strategies for star lattices}
\label{section:design}
\setcounter{figure}{0}   

\begin{figure*}
\begin{center}
\includegraphics[width=0.8\textwidth]{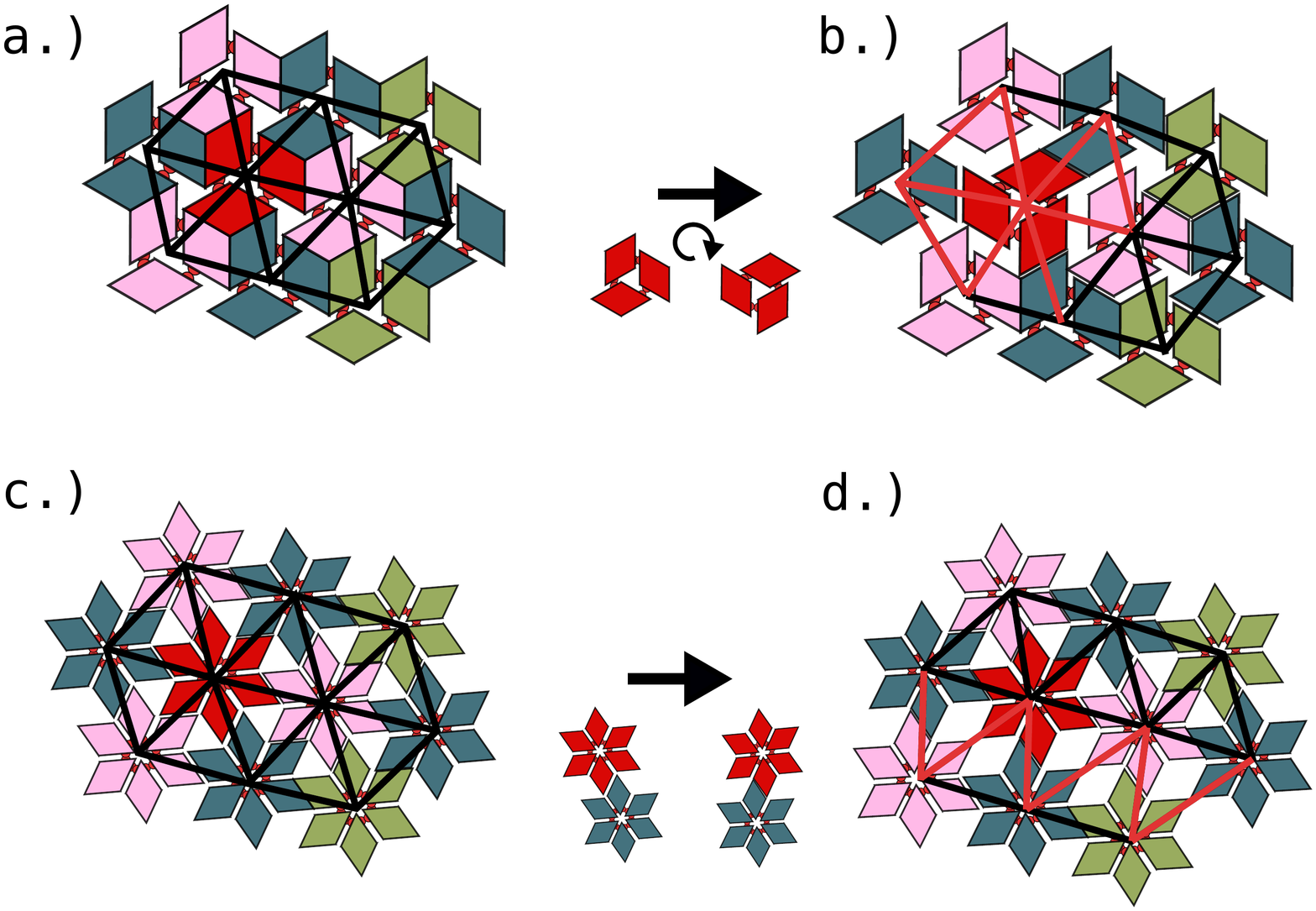}
\caption{Sketches of perfect lattices of open boxes and stars and their respective lattice defects. \textbf{a.)} Open tiling with triangular pores (ot-tiling) composed of open boxes. All open boxes, including the red box are in their correct lattice positions and orientations. The black lines connect the center positions of the open boxes and highlight the hexagonal order of the tiling. \textbf{b.)} The red open box is rotated by $60\degree$ with respect to the perfect ot-tiling in a.). Due to the rotation the rhombi of the red open box are now aligned parallel with their neighbouring rhombi, and hence the center distances changes and the lattice gets distorted. Neighbouring boxes in correct lattice positions are still connected by black lines, while boxes that had to shift due to the orientational defect are now connected with red lines. \textbf{c.)} Open lattice with triangiular pores (ot-tiling) made of 6-stars. All stars, including the red star reside in their correct lattice positions. The hexagonal order of the star centers is highlighted by the black lines connecting the center positions. \textbf{d.)} The red star shifts to align one rhombi on the other edge of its neighbouring rhombi. This results in a line defect where all rhombi in the same line are forced to shift as well. The changed center-to-center distance lines are shown in red, while the lines connecting neighbours that maintain the correct lattice positions are still black.}
\label{fig:defects}
\end{center} 
\end{figure*}

\begin{figure*}
\begin{center}
\includegraphics[width=.8\textwidth]{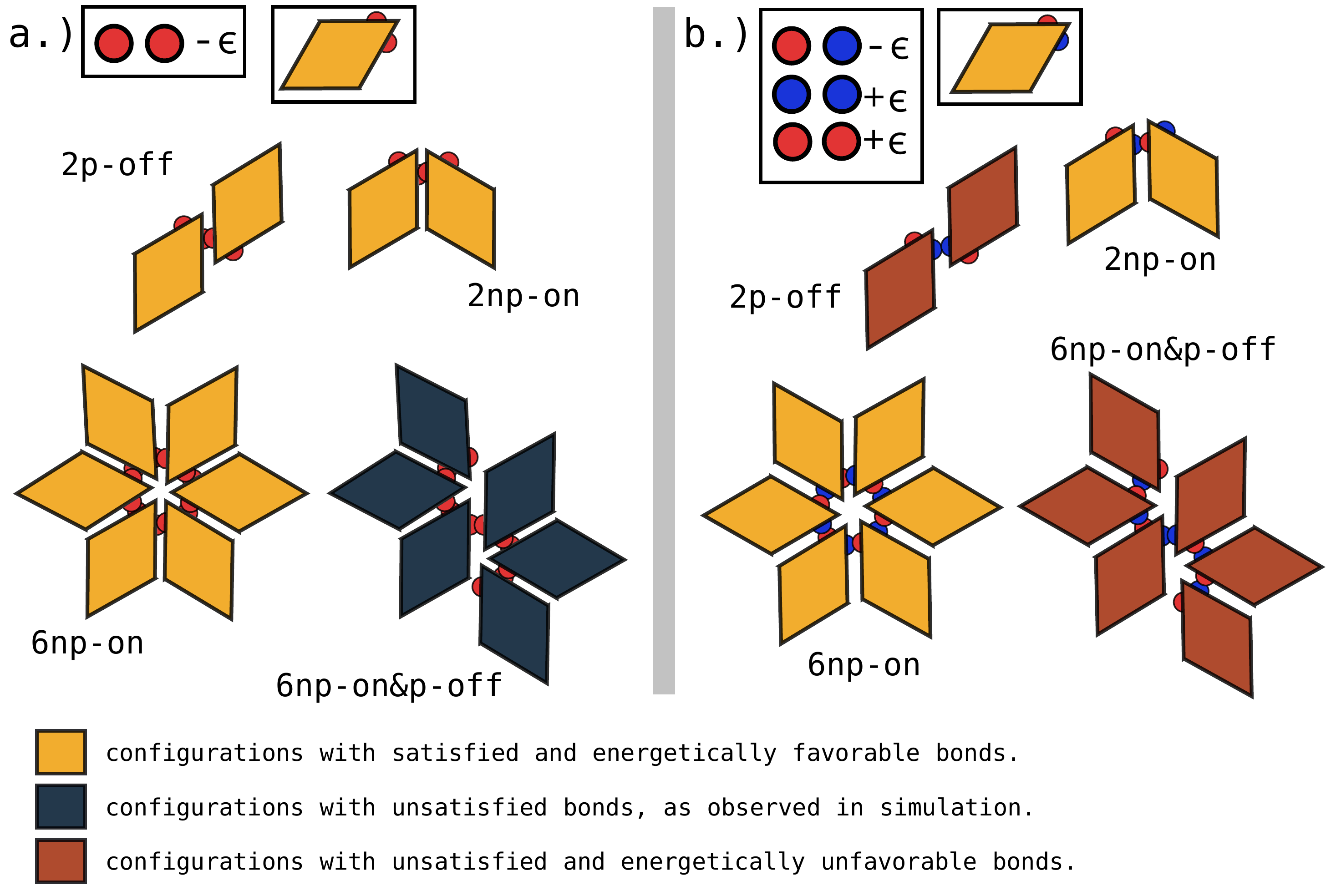}
\caption{Sketches of patch specificities favouring or disfavouring parallel two particle bonds (p-bonds) and subsequently broken stars. \textbf{a.)} Mouse rhombi with attractive patches of the same kind at $\Delta<0.5$, where stars are prevalent. Parallel (2p-off) as well as non parallel (2np-on) bonds are allowed (both in yellow). Stars (6np-on) as well as broken stars with dangling bonds (6np-on$\&$p-off, in dark blue) can form. \textbf{b.)}: Mouse rhombi with two kinds of patches where patches of the same kind repel each other and patches of differing kinds attract each other. Patches are positioned at $\Delta<0.5$, for which stars have been observed. With this specificity 2p-off bonds are not allowed (in burgundy) while 2np-on are still possible (in yellow). This leads to broken stars becoming not allowed (6np-on$\&$p-off, in burgundy) while stars are still allowed (6np-on, in yellow).}
\label{fig:broken_stars}
\end{center} 
\end{figure*}

We observe that in contrast to boxes, stars do not yield an extensive hexagonal lattice at second stage assembly. We identify three factors that lower the star yield and therefore hamper the formation of long range hexagonal order in stars.

Firstly, the highest star yield with $\approx 0.75$ is lower than the yields for boxes and open boxes with over $0.9$. Besides stars, assembly products for mo-s systems contain other clusters that destroy the star lattice order: these clusters are not fully formed stars, p-bonded dimers (labeled as 2p-off in Fig.~\ref{fig:broken_stars}) and broken stars (labeled as 6np-on$\&$p-off in Fig.~\ref{fig:broken_stars}). Broken stars are 5- or 6-particle clusters where parts of the star are shifted and yield parallel and dangling bonds instead of 5 (5-stars) or 6 (6-stars) non parallel bonds (see Fig.~\ref{fig:broken_stars}a). 
Hence one strategy to reduce the yield of p-bonded dimers and broken stars is to generally disfavor p-bonds.

Another effect that lowers the yield might be the competition with chains: we observe that stars only become prevalent in regions of the interaction strength where chains start to compete with stars, $i.e.$, at $\epsilon > 9.2k_{B}T$. Hence, disallowing p-bonds could extend the prevalence region of stars to higher interaction strengths and to $\Delta>0.5$, and in absence of chains higher top yield could be reached. 

Lastly, during compression of the NPT runs, stars, especially 5-stars tend to get destroyed and become broken stars, that in return destroy the long range order. 

Summarizing, the most apparent strategy to get a higher star yield and subsequently long range star lattices is to disfavour p-bonds.
One way to achieve that is to change patch specificities. Instead of one type of patch where patches always attract each other with $-\epsilon$, we propose two types of patches, where patches of the same kind repel each other with $+\epsilon$, while differing patches attract each other with $-\epsilon$.
This scenario disfavors p-bonds and broken stars become unfavorable (see Fig.~\ref{fig:broken_stars}b).
Additionally, as p-bonds are generally disfavored, chains, which need p-bonds, become disfavorable as well. Hence, this might lead to an extended range in the ($\Delta$, $\epsilon$) plane where stars are the dominant self assembly product. 

\clearpage
\footnotesize{
\bibliography{references} 
\bibliographystyle{rsc} 
}
\end{document}